\documentclass[twocolumn,aps,pra,superscriptaddress,nofootinbib,notitlepage,showpacs,floatfix,longbibliography]{revtex4-2}
\usepackage{graphicx} % Required for inserting images
\usepackage{amsmath,amssymb,amsthm,dsfont,physics}
\usepackage[x11names]{xcolor}
\usepackage{mathtools}
\usepackage{braket}
\usepackage{enumitem}
\usepackage{amsthm}
\usepackage{empheq}
\usepackage{subfigure}
\usepackage{cancel}
\usepackage{bm}
\usepackage{mleftright}
\usepackage[normalem]{ulem}
\usepackage{relsize}
\usepackage[pdfstartview=FitH]{hyperref}
\hypersetup{
    colorlinks=true,       % false: boxed links; true: colored links
    linkcolor=blue,          % color of internal links
    citecolor=blue,        % color of links to bibliography
    filecolor=magenta,      % color of file links
    urlcolor=blue,           % color of external links
    runcolor= blue
}
\usepackage[ruled]{algorithm}
\usepackage{algorithmicx}
\usepackage{algpseudocode}
\usepackage{multirow}
\usepackage{booktabs}
\usepackage{hhline}
\usepackage{tabularx}
\usepackage{adjustbox}
\usepackage[margin=1in,a4paper]{geometry}  % choose margins appropriately
\usepackage{array,booktabs,ragged2e}
\usepackage{pifont}
\usepackage{leftindex}
\usepackage{siunitx}
\DeclareSIUnit{\calorie}{cal}
\DeclareSIUnit{\kcal}{\kilo\calorie}
\DeclareSIUnit{\atm}{atm}
\newcolumntype{R}[1]{>{\RaggedLeft\arraybackslash}p{#1}}

\definecolor{light-gray}{gray}{0.95}

\newtheorem*{theorem*}{Theorem}

\newtheoremstyle{problemstyle}  % <name>
        {3pt}                                               % <space above>
        {3pt}                                               % <space below>
        {\normalfont}                               % <body font>
        {}                                                  % <indent amount}
        {\bfseries\itshape}                 % <theorem head font>
        {\normalfont\bfseries:}         % <punctuation after theorem head>
        {.5em}                                          % <space after theorem head>
        {}                                                  % <theorem head spec (can be left empty, meaning `normal')>
\theoremstyle{problemstyle}

\begin{document}
\title{Variational Quantum Annealing for Quantum Chemistry}
\author{Ka-Wa Yip}
\affiliation{Department of Chemistry and Chemical Biology, Northeastern University, Boston, MA 02115, USA}

\author{K\"ubra Yeter-Aydeniz}
\affiliation{Quantum Information Sciences, Optics, and Imaging Department, The MITRE Corporation, 7515 Colshire Drive, McLean, Virginia 22102-7539, USA}
\author{Sijia S. Dong}
\thanks{Corresponding author}
\email{s.dong@northeastern.edu}
\affiliation{Department of Chemistry and Chemical Biology, Northeastern University, Boston, MA 02115, USA}
\affiliation{ 
Department of Physics, Northeastern University, Boston, MA 02115, USA}
\affiliation{ 
Department of Chemical Engineering, Northeastern University, Boston, MA 02115, USA}
\date{\today}
\begin{abstract}
We introduce a variational quantum annealing (VarQA) algorithm for electronic structure theory, in which we use the quantum annealer as a sampler and prepare an ansatz state through its statistics. We also introduce a strategy called the ``digitizer" for searching the space of variational parameters efficiently. We demonstrate the effectiveness of VarQA by evaluating the ground-state potential energy surface
for molecules with up to $20$ spin orbitals as well as an excited-state potential energy surface. This approach resembles the workings of the quantum Boltzmann Machines (QBMs), but is generalized to handle distributions beyond the Boltzmann distribution.
In VarQA, with the number of required logical qubits equal to the number of spin orbitals, a fully connected Ising Hamiltonian can be readily implemented in a large-scale quantum annealer as a scalable ansatz for electronic structure calculations.
\end{abstract}
\maketitle
\section{Introduction}
Quantum annealing (QA)~\cite{Kadowaki1998} is a quantum metaheuristic designed as a quantum-inspired alternative to simulated annealing. By leveraging quantum fluctuations and tunneling, QA allows systems to overcome energy barriers and escape local minima, thereby exploring the solution space more efficiently. QA is part of the broader framework of adiabatic quantum computation (AQC)~\cite{Das2008, Morita2008, Albash2018, Hauke2020, Crosson2021}. Commercial devices, such as those developed by D-Wave Systems, implement QA at the hardware level with over 5,000 qubits~\cite{Dwave,Dwave2024}, making them some of the largest quantum computing systems available. In QA applications, optimization problems are typically formulated using a final Hamiltonian in Ising form, where the ground state represents the solution~\cite{Albash2018}. The adiabatic theorem~\cite{kato1950adiabatic} underpins QA, ensuring that the system evolves toward the solution if the process is sufficiently slow to avoid transitions across energy gaps, thus operating as a genuine time-dependent quantum process.

Accurate evaluation of quantum mechanical properties is crucial in chemistry~\cite{deglmann2015,Bauer2020,PhysRevX.8.011044,doi:10.1021/acs.jctc.4c00544,9781118742631ch03} and physics~\cite{lieb2005stability}, as it provides insights into the fundamental behaviors of molecules and materials. Circuit-based quantum computing has made significant strides in this area, with several gate-based quantum algorithms developed for calculating ground-state energies. These include Quantum Phase Estimation (QPE)~\cite{ADLH05,Lin2022,Dong2022,Wan2022,Ding2023,Wang2023quantumalgorithm,wang2023fastergroundstateenergy}, quantum machine learning~\cite{xia2018quantum,doi:10.1126/science.aag2302}, and the Variational Quantum Eigensolver (VQE)~\cite{Peruzzo2014,kan2017,Tilly2022} with various ansatze, such as unitary Coupled-Cluster (uCC)~\cite{Romero_2019}, qubit coupled cluster singles and doubles (QCCSD)~\cite{xia2020qubit}, and the quantum imaginary time evolution (QITE) algorithm~\cite{gomes2021adaptive, yeter2021benchmarking}. VQE is particularly suited to near-term quantum hardware~\cite{Bauer2020} and has evolved into variants like adaptive VQE (ADAPT-VQE)~\cite{grimsley2019adaptive}, quantum subspace expansion VQE (QSE-VQE)~\cite{colless2018computation}, and multiscale contracted VQE (MC-VQE)~\cite{parrish2019quantum}. However, these VQE algorithms often encounter the barren plateau phenomenon, where the optimization landscape becomes flat due to factors like quantum hardware noise, large circuit size, and excessive entanglement~\cite{ragone2024lie,holmes2022connecting,Singkanipa2025beyondunitalnoisein}.

There has been recent interest in studying electronic structure problems using quantum annealers. A key challenge is mapping the electronic structure Hamiltonian into a form suitable for quantum annealers. The Xia-Bian-Kais (XBK) method~\cite{xia_electronic_2018} addresses this by approximating electronic structure Hamiltonians using an Ising model in an expanded space, allowing the system's ground state energy to be estimated via QA. The Quantum Annealer Eigensolver (QAE)~\cite{tep2019} 
formulates the energy minimization problem as a quadratic unconstrained binary optimization (QUBO) problem over fractionals, with the Lagrange multiplier serving as a variational parameter to ensure the normalization of the fixed-point representation.
This method was tested on the vibrational spectra of molecules including O$_2$ and O$_3$ and later generalized to complex Hermitian matrices and electronic structures~\cite{teplukhin2020electronic}. 
Recent interest has also emerged in operating quantum annealers in a variational manner. Variational Coherent Quantum Annealing (VCQA)~\cite{barraza2023variational} seeks to enlarge the annealing gap by variationally modifying the annealing schedule and introducing a third Hamiltonian. This, in turn, extends the coherence time of the quantum annealer. However, VCQA is unrelated to quantum chemistry or electronic structure problems.

In this work, we introduce a variational quantum annealing (VarQA) algorithm that addresses the limitations of existing algorithms for solving eigenvalue problems on quantum annealers.  
Unlike previous methods --- XBK and QAE --- VarQA is a fully variational algorithm where the annealer functions as a thermal sampler. It does not expand over the Hilbert space of the electronic Hamiltonian, requiring a number of logical qubits equal to the number of (spin) orbitals in a molecule.
Theoretically, adiabatic quantum computing is equivalent to circuit-based quantum computing and can be equally powerful~\cite{aharonov2008}. We aim to adapt the successful methodologies of VQE for ground state evaluation to the quantum annealing framework, which offers advantages such as more qubits and robustness to noise and gate errors~\cite{Preskill_2018}.

This paper is organized as follows. In Sec.~\ref{sec:background}, we provide an overview of the key concepts used in this work, including the electronic structure Hamiltonian (Sec.~\ref{sec:elec_struc}) and the Ising Hamiltonian (Sec.~\ref{sec:ising}) that is utilized in quantum annealer. We also discuss the quantum annealing process (Sec.~\ref{sec:qa}), and how it operates in an open system (Sec.~\ref{sec:qa_open}). We then present the details of the proposed VarQA algorithm in Sec.~\ref{sec:method}, along with its key component: parameterized trial state construction. Additionally, we introduce a digitization technique of the Ising energy landscape. In Sec.~\ref{sec:results}, we present our findings on calculating ground and excited state energies for H$_2$ and ground state energies for LiH, He$_2$, HeH$^{+}$ and O$_2$ using VarQA algorithm on D-Wave quantum annealer as well as the values obtained using simulated annealing method, and compare them to the exact values obtained using exact diagonalization. In Sec.~\ref{sec:methodcomparison}, we analyze the qubit resource requirements for the VarQA algorithm in relation to other quantum annealing methods. We finalize the paper with Sec.~\ref{sec:conclusion} where we discuss our findings and present path forward studying larger molecules using VarQA algorithm.

\section{Background}
\label{sec:background}
In this section, we consider two Hamiltonians to study ground and excited state energies of the molecules of interest using the VarQA algorithm. The first is the electronic Hamiltonian, which serves as the target, and the second is the Ising Hamiltonian. Our objective is to use the Ising Hamiltonian as an ansatz to construct a trial state that approximates the ground state of the electronic Hamiltonian.

\subsection{Electronic Structure Hamiltonian}
\label{sec:elec_struc}
We first focus on the electronic structure Hamiltonian of a molecule in the second-quantized form~\cite{Helgaker_ch1,szabo1996modern}. Given a set  
of $M$ (spin) orbitals $\{\phi_1(\bm{x}),\dots,\phi_M(\bm{x})\}$, where $\bm{x}$ represents the combined spatial and spin coordinates of the electron,
the electronic Hamiltonian $H_{\text{elec}}$ is given by:
\begin{align}
\label{eq:fermionic}
H_{\text{elec}} 
&=\sum_{p,q=1}^{M} h_{pq} 
a^\dag_{p} a_{q} +\frac{1}{2}\sum_{p,q,r,s=1}^{M} g_{pqrs}a^\dag_{p} a^\dag_{q} a_{r} a_{s} \,,
\end{align}
where $a_{p}$ and $a^{\dagger}_{p}$ are the lowering and raising operators for orbital $p$ respectively, satisfying the anticommutation condition $\{a^\dagger_p, a_q\} = \delta_{pq}$. 
In atomic units, 
\begin{align}
    h_{pq} &= \int  \phi_p^* \left(\bm{x}\right) \left(-\frac{\nabla^2}{2} - \sum_{A}\frac{Z_{A}}{r_{A}}\right) \phi_q \left(\bm{x}\right) d\bm{x}\,,\\
    g_{pqrs} &= \iint \frac{\phi_p^*(\bm{x}_1) \phi_q^*(\bm{x}_2) \phi_r(\bm{x}_2)  \phi_s(\bm{x}_1)}{r_{12}} \,d\bm{x}_1d\bm{x}_2\,,
\end{align}
where $Z_A$ represents the nuclear point charges, $r_A$ the electron-nuclear distances, $r_{12}$ the electron-electron distance.

For the purpose of quantum computation in qubit form, we perform the Jordan-Wigner transformation~\cite{jordan_ber_1928, aspuru2005} to the second-quantized Hamiltonian. Simply put, the following substitutions are applied in Hamiltonian \eqref{eq:fermionic} :
\begin{equation*}
    \hat{a}_j \mapsto \frac{1}{2}(\sigma^x_j+i\sigma^y_j) \sigma^z_{1} \cdots \sigma^z_{j-1} \,.
\end{equation*}
The result is an $M$-qubit Hamiltonian: 
\begin{equation}
\label{eq:pauliform}
    H_{\text{elec}} = \sum_{i}\gamma_iP_i \,,
\end{equation}
with $\gamma_i \in \mathbb{R}$ and the Pauli bases $P_i = \bigotimes_{j=1}^{M}\sigma_{i,j}$ for $\sigma_{i,j} \in \{I, \sigma^x, \sigma^y, \sigma^z\}$. Eq.~\ref{eq:pauliform} is an exact representation of the original second-quantized Hamiltonian in Eq.~\ref{eq:fermionic}.

In quantum chemistry, calculating the ground state, $\ket{\psi_{g}}$, and the corresponding ground state energy $E_{g}$ of the electronic structure Hamiltonian of a molecule is often of significant interest, which are defined as follows:
\begin{equation}
\label{eq:gs}
\langle\psi_{g}|H_{\text{elec}}|\psi_{g}\rangle = E_{g}
\end{equation}
One approach to solve this equation is to utilize exact diagonalization (ED). However, ED is computationally intractable for larger molecular systems due to exponential growth of the Hilbert space. Hence, a more practical method is to employ the variational principle, which provides an upper bound to the ground state energy and is defined as
\begin{equation}
\label{eq:vp}
\min_{\bm{\theta}}\langle\psi(\bm{\theta})|H_{\text{elec}}|\psi(\bm{\theta})\rangle \geq E_{g} \,,
\end{equation}
such that at the optimal parameters 
$\bm{\theta}^{\ast}$, $\langle\psi(\bm{\theta}^{\ast})|H_{\text{elec}}|\psi(\bm{\theta}^{\ast})\rangle$ approaches %\gtrapprox 
$E_g$. 
The exact choice of variational parameters $\bm{\theta}$ is given in Sec.~\ref{sec:ising}. The variational principle is also the foundation of our proposed VarQA algorithm. 

\subsection{Ising Hamiltonian}
\label{sec:ising}
Our proposed VarQA algorithm is designed for quantum annealers, which natively solve Ising models. Hence, next, we consider a quadratic Ising Hamiltonian without the transverse field. For $M$ qubits, it is given by:
\begin{align}
\label{eq:isingHamiltonian}
H_{\text{Ising}} &= \sum_{i=1}^{M}h_{i}\sigma_{i}^{z} + \sum_{j>i=1}^{M}J_{ij}\sigma_{i}^{z}\sigma_{j}^{z} \,.
\end{align}

In this work, $h_i$, i.e., longitudinal fields applied to each qubit and $J_{ij}$, i.e., interaction couplings between qubits, serve as the variational parameters. An identity term can also be added to the Ising Hamiltonian as an overall offset. With this addition, there are a total of $\nu = M(M+1)/2+1$ degrees of variational freedom, and we have the parameterized Hamiltonian as
\begin{align}
\label{eq:parameterizedHamiltonian}
H(\bm{\theta}) &= \sum_{i=1}^{M}\theta_{i}\sigma_{i}^{z} + \sum_{j>i=1}^{M}\theta_{ij}\sigma_{i}^{z}\sigma_{j}^{z} + \theta_0 I \,,
\end{align}
with variational parameters (angles) 
$\bm{\theta} = \{\theta_0,\bm{\theta}_{i}, \bm{\theta}_{ij}\} \in \mathbb{R}^{\nu}$. The next sections (Sec.~\ref{sec:qa}-~\ref{sec:finalstate} and~\ref{sec:varqa}) describes how to prepare the parameterized trial state 
$\ket{\psi(\bm{\theta})}$ from $H(\bm{\theta}) $.

\subsection{Quantum annealing (QA)}
\label{sec:qa}
Taking into account the variational angles $\bm{\theta}$ of the parameterized Hamiltonian,  the QA is governed by the following time-dependent Hamiltonian:
\begin{equation}
\label{eq:annealH}
    H(t, \bm{\theta}) = -\frac{A(t)}{2} \left(\sum_{i=1}^M \sigma_i^x\right)+ \frac{B(t)}{2} H(\bm{\theta})\,.
\end{equation}
The first term in \eqref{eq:annealH} represents the initial Hamiltonian of the system, and the second term is the final Hamiltonian of the system. $A$ and $B$ are the coefficients that scale the transverse field ($\sigma_i^x$) and Ising contributions, respectively. Starting at $t=0$,  the system begins with a sum of transverse fields and undergoes adiabatic evolution, ending at  $t=t_f$ with the Hamiltonian $H(\bm{\theta})$\footnote{Note that it is scaled with a constant $B(t_f)/2$, which we omit for simplicity in the rest of the analysis.}. In this context, the total anneal duration is $t_f$. If the annealing is slow enough the system remains in its lowest energy state and at the end of annealing it reaches the ground state of the problem Hamiltonian.

The initial state of QA is the ground state of the sum of transverse fields, i.e., 
\begin{equation*}
    \ket{\psi(0)} = \bigotimes_{i=1}^{M}\ket{+}\,,\text{\,\,\,\,\,\,where\,\,\,\,\,\,\,} \ket{+} = \frac{\ket{0}+\ket{1}}{\sqrt{2}} \,.
\end{equation*}

\begin{figure}[htb!]
    \centering
\includegraphics[width=0.43\textwidth]{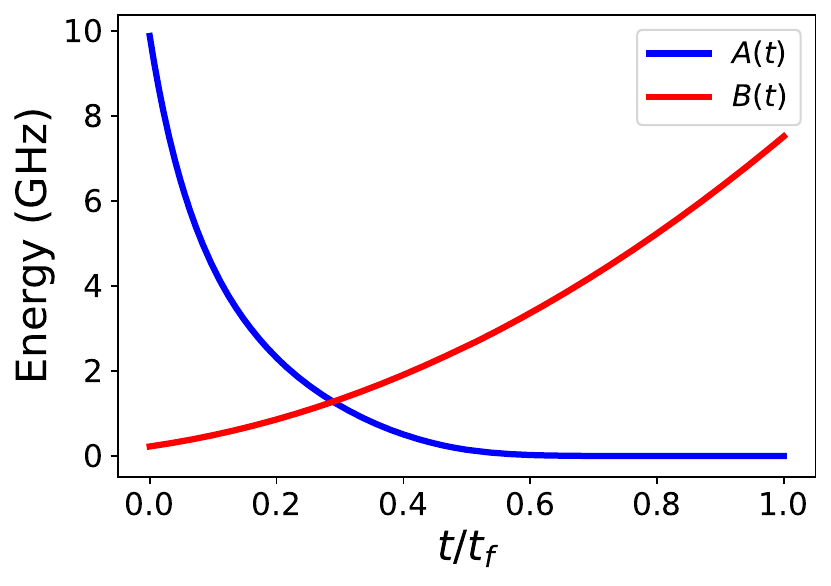}
    \caption{Annealing schedule of D-Wave's Advantage 2 Prototype 7.1 QPU~\cite{dwavespec}. The total anneal time is denoted by $t_f$. 
    }
    \label{fig:d-wave_schedules}
\end{figure}

D-Wave machines have some particular annealing schedules (e.g., see Fig.~\ref{fig:d-wave_schedules}). The nonlinear features of such schedules are crucial for the production of a high-quality Gibbs state~\cite{qbm}.

\subsection{Quantum annealing in an open system}
\label{sec:qa_open}
QA hardware nonetheless operates in an open-system environment subject to thermal noise. Instead of a pure state, a quantum annealer produces a mixed state $\rho(t_f,\bm{\theta})$
at the end of the annealing process, i.e.,
\begin{equation}
\rho(t_f,\bm{\theta})=\Phi(t_f, \bm{\theta})[\rho(0)]\ ,
\end{equation}
where $\rho(0) = \ketbra{\psi(0)}{\psi(0)}$ represents the initial state of the quantum system, and $\Phi(t_f, \bm{\theta})$ is a quantum channel that acts over a duration $t_f$ and depends on the variational angles $\bm{\theta}$.

In an open system, the evolution of the system is governed by a Lindblad master equation, which describes the non-unitary evolution of the system due to interactions with the environment. If the evolution is Lindbladian (i.e., it follows Lindblad master equation), then the quantum channel $\Phi(t_f, \bm{\theta})$ can be represented in the following form:
\begin{equation}
\label{eq:lindbladian_channel}
\Phi(t_f, \bm{\theta})= {\cal T}\exp\left[\int_{0}^{t_f}\mathcal{L}(t,\bm{\theta})dt\right]\ ,
\end{equation}
where $\cal T$ denotes time order and $\mathcal{L}(t,\bm{\theta})$ denotes the time-dependent Liouvillian superoperator, which governs the system's open dynamics. For the example of a master equation~\cite{Albash2012,Yip2018}, the Liouvillian, neglecting the Lamb shift, is given by:
\begin{equation}
    \mathcal{L}(t,\bm{\theta}) = -i\left[H(t,\bm{\theta}),\boldsymbol{\cdot}\right]+\mathcal{D}(t,\bm{\theta})\,,
\end{equation}
where $\mathcal{D}(t,\bm{\theta})$ is the dissipator in the weak coupling limit that also depends the exact angles $\boldsymbol{\theta}$.

The main implication, here, is that if the anneal time $t_f$ is long enough, the quantum annealer can reach the steady state of a Gibbs distribution. Specifically, as $t=t_f\rightarrow \infty$, the steady state system satisfies $\mathcal{L}(t_f,\bm{\theta})\rho_{\textrm{SS}}(t_f,\bm{\theta}) = 0$, with 
\begin{equation}
\label{eq:Gibbs_state}
\rho_{\textrm{SS}}(t_f, \bm{\theta}) = \frac{e^{-\beta H(\bm{\theta})}}{\text{Tr}(e^{-\beta H(\bm{\theta})})}\,.
\end{equation}
The denominator is the partition function and $\beta = 1/T$ represents the inverse temperature, assuming that the Boltzmann constant, $k_B$, is set to $1$. For D-Wave's quantum processing unit (QPU)~\cite{Dwave2024}, the (effective) temperature $T$ is approximately $1.56$ GHz (or $12$ mK). The $t_f$ value needed for the system to equilibrate into the Gibbs state is detailed in~\cite{PhysRevA.95.042302,PhysRevApplied.19.034095}, which, in our case, also depends on the variational angles $\bm{\theta}$. 

\subsection{Final state of quantum annealing}
\label{sec:finalstate}
Overall, the final state at the end of the anneal is a thermal state $\rho(t_f,\bm{\theta})$. Given that it is a Gibbs state, one could obtain a Boltzmann distribution (with an effective temperature) from a sample of measurements. However,  $\rho(t_f,\bm{\theta})$ is often not an exact Gibbs state due to several factors: 
\begin{itemize}
    \item The total anneal time $t_f$ is not long enough for the system to equilibrate into the steady state;
    \item The quantum channel $\Phi$ is not Lindbladian~\cite{PhysRevApplied.17.054033};
    \item Additional system effects, such as freeze-out phases, control imperfections, and electronic errors~\cite{PhysRevApplied.17.044046}.
\end{itemize}

Nevertheless, as discussed in the following section, our algorithm does not require the final state $\rho(t_f,\bm{\theta})$ to be a Gibbs state. We present its inner workings, with the Gibbs state as a special case.

\section{Method}
\label{sec:method}
The VarQA algorithm is designed to study electronic structures 
by leveraging
the principles of quantum annealing. It defines the ansatz as a fully connected Ising Hamiltonian, with variational parameters being the strengths of transverse fields and qubit couplings. By tuning these parameters, we can influence the annealing gap size and spectrum, thereby shaping the final quantum state distribution.

While the adiabatic theorem suggests the system evolves to the ground state, quantum noise and decoherence are inevitable in Noisy Intermediate-Scale Quantum (NISQ) technology~\cite{Preskill_2018}. Thermal excitations around the minimum gap point~\cite{Albash2012,Yip2018} can lead to a mixed final state. If thermalization is sufficient, the final state may approach a Gibbs state~\cite{PhysRevA.95.042302}, 
though this is not always achievable due to the factors mentioned in Sec.~\ref{sec:finalstate}. 
However, techniques exist to produce high-quality Gibbs states in annealers, especially for certain problem classes~\cite{PhysRevApplied.17.044046}.

Our VarQA algorithm starts with a parameterized trial state preparation. To construct the trial state for the variational principle~\cite{Yourgrau1968-YOUVPI-6}, we gather statistics over a sample of annealing processes. If the final state resembles a Gibbs state, the statistics may follow a Boltzmann distribution. The theory of Quantum Boltzmann Machines (QBMs)~\cite{qbm} has been explored in machine learning~\cite{Amin2018,Benedetti2017,Kieferova2017}, and using QBM as an ansatz avoids the barren-plateau problem~\cite{Coopmans2024}. Recent work has shown that such a variational approach is sample-efficient for ground-state energy estimation~\cite{patel_quantum_2024}. Our approach extends beyond Boltzmann statistics to other open system statistics, enabling the algorithm to work with real noise statistics in commercial quantum annealers, potentially without quantum error correction.

To efficiently prepare the parameterized trial state for VarQA, we introduce a ``digitizer" technique to exploit the structure of the Ising energy landscape. By focusing on the parameter space of $h_i, J_{ij} \in \{-1,0,+1\}$ instead of that consists of real numbers, we control the degeneracy levels of the Ising spectrum, targeting sparse Hamiltonians. This reduces the search space for optimal parameters, and in many cases, the ``digitizer" alone provides a highly accurate solution, minimizing the need for further tuning. In addition, the ``digitizer" allows us to leverage parallel computing to significantly accelerate the search for optimal parameters.

\subsection{Variational Quantum Annealing (VarQA)}
\label{sec:varqa}
Here we present the components of our VarQA algorithms: 1. the construction of a parameterized trial state, 2. an optimization technique called the ``digitizer", and 3. an overview of the algorithm.

\subsubsection{Parameterized trial state construction}
\label{sec:parameterized_construction}
Here we present how to construct the parameterized trial state   
$\ket{\psi(\bm{\theta})}$ from the final state $\rho(t_f,\bm{\theta})$.
At the end of the anneal, the system evolves into a mixed final state $\rho(t_f,\bm{\theta})$. To obtain its information, one can perform quantum measurements with measurement operators $\{A_{m}\}$. Here $A^{\dagger}_{m}A_{m}$ is positive with 
$\sum_{m}A^{\dagger}_{m}A_{m} = I$, and $m$ is the measurement outcome (bit string) corresponding to the measurement operator $A_{m}$. The probability of obtaining the measurement outcome $m$ is
\begin{equation}
\label{eq:p_1}
    p(m) = \Tr(A^{\dagger}_{m}A_{m}\rho(t_f,\bm{\theta})) \,.
\end{equation}
One can construct the trial state from an $M$-qubit final state $\rho(t_f,\bm{\theta})$ as a coherent superposition of the measurement outcomes:
\begin{equation}
    \label{eq:trial_state}
    \ket{\psi(\bm{\theta})}_{\pm} = \sum_{m}\varepsilon_m \sqrt{p(m)}\ket{m},\,\,\text{where  }\varepsilon_m=\pm 1\,,
\end{equation}
for every possible combination of signs $\pm$ inside the summation. The specific combinations of signs $\pm$ may be constrained by sign rules, such as the Marshall-Peierls sign rule~\cite{marshall1955}. 

If $\ket{m}$ is in the Ising (or computational) basis, then   
Eq.~\ref{eq:p_1}
reduces to 
\begin{equation}
    \label{eq:p_2}
    p(m)= \langle m|\rho(t_f,\bm{\theta})|m\rangle\,,
\end{equation}
since the diagonal elements of $\rho$ are the  probabilities from the computational basis measurement.

Furthermore, if $\rho(t_f,\bm{\theta})$ is a Gibbs state, i.e.,
\begin{equation*}
    \rho(t_f,\bm{\theta}) = \frac{e^{-\beta H(\bm{\theta})}}{\text{Tr}(e^{-\beta H(\bm{\theta})})}\,, 
\end{equation*}
it is inherently diagonal in the computational basis. In this case, $\rho(t_f,\bm{\theta})$ is also the QBM state in various literatures~\cite{patel_quantum_2024,Zoufal_2021,huijgen2024}. Since the parameterized Hamiltonian $H(\bm{\theta})$ is an Ising Hamiltonian, 
Eq.~\ref{eq:p_1} 
can be naturally expressed as
\begin{equation}
    \label{eq:p_3}
    p(m) =     \frac{e^{-\beta E_m(\bm{\theta})}}{\sum_{m}e^{-\beta E_m(\bm{\theta})}}\,,
\end{equation}
where $E_{m}(\bm{\theta})$ is from
\begin{equation}
    H(\bm{\theta}) = \sum_{m}E_{m}(\bm{\theta})|m\rangle\langle m| \,,
\end{equation}
which is the spectral decomposition of  $H(\bm{\theta})$.
Here $E_{m}(\bm{\theta})$ stands for the eigenenergies and is a function of the parameterized angles $\bm{\theta}$.

\subsubsection{Digitizing the energy landscape --- the ``digitizer"}
\label{sec:digitizer}
For a given $M$-orbital electronic Hamiltonian $H_{\text{elec}}$ and its ground state $\ket{\psi_g}$, we want to assemble a $\ket{\psi(\bm{\theta})}$, using Eq.~\ref{eq:trial_state}, to solve the following minimization program:
\begin{align}
\label{eq:min_program}
&\min_{\bm{\theta},\,\pm}\,
{}_{\pm}\langle\psi(\bm{\theta})|H_{\text{elec}}|\psi(\bm{\theta})\rangle_{\pm}  \notag\\
     &\,\,\text{s.t.} \,\,\, \bm{\theta} \in \mathbb{R}^{\nu}, \,\,\nu = M(M+1)/2+1 \,.
\end{align}
While the number of variational parameters, $\nu$, scales polynomially with the number of spin orbitals (or qubits) $M$ in the system, Eq.~\ref{eq:trial_state} contains exponentially many terms, resulting in an exponential number of possible sign combinations. For our purposes, the second-quantized electronic Hamiltonian $H_{\text{elec}}$ is highly sparse~\cite{PhysRevA.104.042607}, especially for large systems, --- they typically consist of a number of terms polynomial with respect to the problem size, with each row or column containing only a polynomial number of nonzero entries. 
Therefore, the number of terms for $\ket{\psi(\bm{\theta})}$ in the relevant parameter space is also polynomial in $M$, and the optimal signs can be efficiently determined for each set of parameterized angles $\bm{\theta}$ (See also Algorithm~\ref{alg:1}).

The exact number of (nonzero) terms in Eq.~\ref{eq:trial_state} primarily depends on the level of degeneracy and the energy gap relative to the lowest energy state in the parameterized Hamiltonian $H(\bm{\theta})$.
In general, the higher the degeneracies and the larger the gaps to the ground state, the fewer terms appear in $\ket{\psi(\bm{\theta})}$. 
To efficiently create energy degeneracies and change energy gaps, we introduce a ``digitizer" technique, which effectively discretizes the continous energy landscape of the Ising Hamiltonian $H(\bm{\theta})$. 

First we focus our choices of variational parameters on the following two sets, $\mathcal{D}_1$, $\mathcal{D}_2$, of integer parameters 
% \kya{I think this should be in the text not as a footnote} 
(as shown in~\cite{PhysRevApplied.17.044046}, such choices also lead to the production of high-quality Gibbs states in the D-Wave annealer):
\begin{align}
    \mathcal{D}_1 &= \{-1,1\}^{\otimes \nu}\,,\\
    \mathcal{D}_2 &= \{-1,0,1\}^{\otimes \nu}\,.
\end{align}
Since all the multi-qubit Pauli operators of the Ising Hamiltonian consist only of $\sigma_{i,j} \in \{I, \sigma^z\}$, selecting $\bm{\theta}$ in either $\mathcal{D}_1$ or $\mathcal{D}_2$ can yield a highly degenerate energy spectrum for $H(\bm{\theta})$. This choice also results in energy gaps to the ground state that are large enough to produce localized measurement outcomes --- leading to a sparse coherent superposition in Eq.~\ref{eq:trial_state}. After pinning down the locations of nonzero terms, one can further fine-tune the values of $\bm{\theta}$ to improve accuracy. We elaborate on these intricacies in Appendix~\ref{app:digital}.

\subsubsection{Outline of the VarQA algorithm}
For a single anneal, where $t$ evolves from $0$ to $t_f$, one measurement yields a specific bit string outcome. To obtain a statistical sample of output distributions, we perform $S$  anneals, where $S$ represents the sample size. These output statistics are then used to construct a trial ground state (Eq.~\ref{eq:trial_state}). 
With $S$ samples of measurement outcomes $\{m^s\}$, one can estimate the probability of each bit string $m$ appearing and use these probabilities to construct a parameterized trial state as follows:
\begin{equation}   
\label{eq:trial_state_exp}
    \ket{\psi(\bm{\theta})}_{\pm} = \sum_{m}\varepsilon_m \sqrt{\frac{\sum_{s=1}^{S}[m^s=m]}{S}}\ket{m},\,\varepsilon_m=\pm 1\,.
\end{equation}
In Eq.~\ref{eq:trial_state_exp}, we count the ratio of measurement outcomes $\{m^s\}$ that correspond to each specific bit string $m$.

With this trial wavefunction, formed from Ising (or computational) basis sampling, we can efficiently compute the expected energy using classical methods~\cite{PhysRevResearch.4.033173}. Given the computational-basis measurement statistics, which determine the exact form of the trial wavefunction, the evaluation of ${}_{\pm}\langle\psi(\bm{\theta})|H_{\text{elec}}|\psi(\bm{\theta})\rangle_{\pm}$ consists of computing a series of transition element $\langle m|H_{\text{elec}}|n\rangle$, where $p(m), p(n) \neq 0$. 
This calculation is algebraic and algorithmically efficient, as the computation time scales linearly with both the number of qubits and the number of Pauli strings in the Hamiltonian (Eq.~\ref{eq:pauliform}). We provide a more detailed elaboration on this in Appendix~\ref{app:exp_eval}. Through a series of ansatz updates, the lowest expected energy,
$\langle\psi(\bm{\theta}^{*})|H_{\text{elec}}|\psi(\bm{\theta}^{*})\rangle$, is selected, corresponding to the optimal ansatz $\bm{\theta}^{*}$.

\begin{figure*}[htb!]
    \centering
\includegraphics[width=1.99\columnwidth]{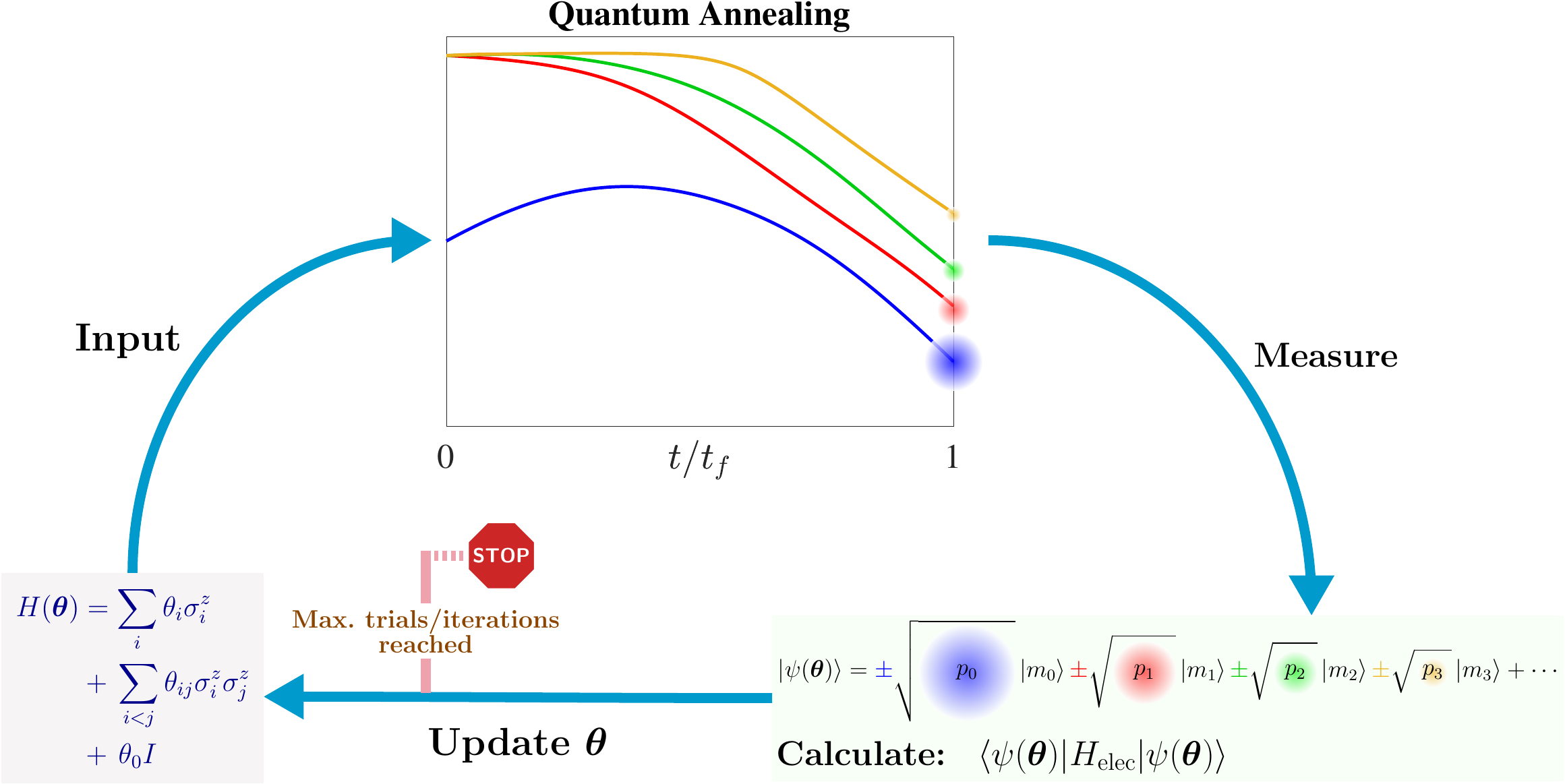}
    \caption{A schematic overview of the VarQA algorithm: quantum annealing, parameterized trial state preparation from a sample of measurements, and ansatz update. The search for the variational angle $\bm{\theta}$ does not need to be iterative and can be parallelized when our digitizer technique is used.}
    \label{fig:schematics}
\end{figure*}

Fig.~\ref{fig:schematics} presents a schematic overview of the VarQA algorithm, illustrating the three main components in this hybrid quantum-classical framework. Algorithm~\ref{alg:1} provides an abstract description of the VarQA algorithm, that searches over all possible solutions (``brute-force"). In Algorithm~\ref{alg:1}, the outermost for-loop spans an infinite number of variational angles. In practice, for a finite number $T$ of variational parameter sets, the VarQA algorithm requires $S \times T$ anneals. 
All these anneals can in principle be executed in parallel~\cite{pelofske_parallel_2022} which can significantly reduce the algorithm runtime. In implementation, $T$ depends on the specific optimization techniques used in the VarQA algorithm. For example, it may represent the size of the digitizer ($|\mathcal{D}_1|$ or $|\mathcal{D}_2|$), or the maximum number of iterations in a gradient descent optimization.

\begin{algorithm}[H]
\caption{Variational quantum annealing (\textit{brute-force})}
\textbf{Input}: An electronic Hamiltonian $H_{\text{elec}}$ of $M$ orbitals 
\begin{algorithmic}[1]
\State Choose a number of samples $S$.
\Procedure{VarQA}{}
\For{every $\bm{\theta} \in \mathbb{R}^{\nu}$} \Comment{The outer loop can be replaced with a finite set of angles, e.g., $\mathcal{D}_1$ and $\mathcal{D}_2$.}
\For{$s=1$ to $S$}
	 	\State Perform a forward anneal with $H(\bm{\theta})$. 
	 	\State Perform measurement in the computational basis.
\EndFor
\State Obtain the bit-string distribution of $S$ 
samples.
\For{every possible sign $\pm$}
\State Reconstruct a state vector $\ket{\psi(\bm{\theta})}_{\pm}$
\State Evaluate 
${}_{\pm}\langle\psi(\bm{\theta})|H_{\text{elec}}|\psi(\bm{\theta})\rangle_{\pm}$
\EndFor
\State Set $\langle\psi(\bm{\theta})|H_{\text{elec}}|\psi(\bm{\theta})\rangle = \min_{\pm}\{{}_{\pm}\langle\psi(\bm{\theta})|H_{\text{elec}}|\psi(\bm{\theta})\rangle_{\pm}\}$
\EndFor
\State Angle searching: Search for the angles
$\bm{\theta}^{*}$,
that minimize: $\langle\psi(\bm{\theta})|H_{\text{elec}}|\psi(\bm{\theta})\rangle$.   
\EndProcedure
\State{\Return $\theta^{*}, \ket{\psi(\bm{\theta}^{*})}, \langle\psi(\bm{\theta}^{*})|H_{\text{elec}}|\psi(\bm{\theta}^{*})\rangle$.}
\end{algorithmic}
\label{alg:1}
\end{algorithm}

\section{Results}
\label{sec:results}
We tested the VarQA algorithm for the ground and excited state energy calculations of several molecules including $\mathrm{H_2}$, $\mathrm{He_2}$, $\mathrm{HeH^+}$,  $\mathrm{LiH}$, and $\mathrm{O_{2}}$. We used the STO-3G basis set for all molecules. For $\mathrm{H_2}$, $\mathrm{He_2}$, and $\mathrm{HeH^+}$, we used $M=4$ (spin) orbitals, requiring $4$ qubits. For $\mathrm{LiH}$, we consider three cases: (1) with only $2$ active electrons and $4$ active orbitals, requiring $4$ qubits; (2) with the innermost $2$ orbitals occupied by $2$ electrons and frozen, requiring $10$ qubits; and (3) with no restrictions on the $4$ electrons and all $12$ orbitals, requiring $12$ qubits. To distinguish the first two cases from the original LiH molecule, we designate (1) as ``active-space LiH" ($\mathrm{as}\text{-}\mathrm{LiH}$) and (2) as ``frozen-core LiH" ($\mathrm{fc}\text{-}\mathrm{LiH}$). For $\mathrm{O_{2}}$, no restrictions were imposed, requiring $20$ qubits. The molecules' second-quantized Hamiltonians and the actual ground state energies were prepared and calculated using IBM Qiskit~\cite{javadiabhari2024quantumcomputingqiskit}, which utilizes the PySCF package~\cite{pyscf}. For optimal efficiency, we also utilize PyTorch Tensors~\cite{torch} for matrix-vector manipulations.

Due to limited access privileges to the D-Wave QPUs, we did not perform QA for all cases. Instead, we performed most of the calculations using Simulated Annealing (SA)~\cite{metropolis1953, bertsimas1993simulated,ackley},  through the D-Wave's Ocean tools. SA is used here as a preliminary demonstration. The final goal is to use QA, which offers greater speed and scalability for large problems, leveraging quantum phenomena such as quantum tunneling through energy barriers~\cite{Albash2018}. Note that when using SA, the statistics in Eq.~\ref{eq:trial_state_exp} converge to a Boltzmann distribution, while this is not necessarily the case for QA. 
The SA results are generated in full parallelism with the digitizer technique. 

As mentioned earlier, for smaller molecules, including $\mathrm{H_2}$, $\mathrm{He_2}$, $\mathrm{HeH^+}$, as well as in the case of $\mathrm{as}\text{-}\mathrm{LiH}$, we mapped the problem onto four qubits using $M=4$ spin orbitals. This type of mapping requires $\nu=11$ variational parameters, hence considering all possible angles leads to trial sizes of $T=|\mathcal{D}_1|=2^{11}$ and $T=|\mathcal{D}_2|=3^{11}$ for digitizers $\mathcal{D}_1$ and $\mathcal{D}_2$, respectively. For larger systems, including $\mathrm{fc}\text{-}\mathrm{LiH}$, LiH, and O$_2$, the number of variational parameters is $\nu=56$, $\nu=79$, and $\nu=211$, respectively. Therefore, for these molecules we focus on the digitizer $\mathcal{D}_1$. Rather than exploring all possible angles in $\mathcal{D}_1$, we randomly select a variational angle set from $\mathcal{D}_1$ in each trial. The number of random trials is $T=100,000$. For details on how the number of random trials affects the VarQA results, please refer to Appendix~\ref{app:trials_samples}.
We also chose the number of samples to be $S=1000$ for all the molecules studied. We computed the ground-state bond dissociation curve of each of the molecules and the first excited state of $\mathrm{H_2}$ using VarQA. The energies computed using VarQA were compared with the exact results, obtained by direct diagonalization.

\subsubsection{Ground state evaluation}
For the ground state energy calculations of these molecules using the VarQA algorithm, we utilized both SA and QA and compared their performances. The QA experiments were conducted on D-Wave's Advantage 2 quantum processing unit (QPU) using the annealing schedules shown in Fig.~\ref{fig:d-wave_schedules}. Because of limited access to D-Wave QPU resources, we executed the QA process serially as follows.  
First, we used the optimal integer parameters of $\mathcal{D}_1$ determined by SA as warm-start parameters for QA. 
We then refine these integer parameters, extending them beyond discrete values, to assess whether a lower expectation value of energy can be achieved.
Note that due to limited access to QPU, such refinements were performed only on interatomic distances where QA and SA results exhibit noticeable deviations; we could also only randomly perturb a limited number of integer parameters and selected the one with the lowest expected energy. Due to noise fluctuations in the quantum hardware, even for a fixed set of parameters, we ran the same QA experiments three times to select the one with the lowest energy. 

\begin{figure*}[htb!]  
  \subfigure[]{
  \centering
\includegraphics[width=1.43\columnwidth]{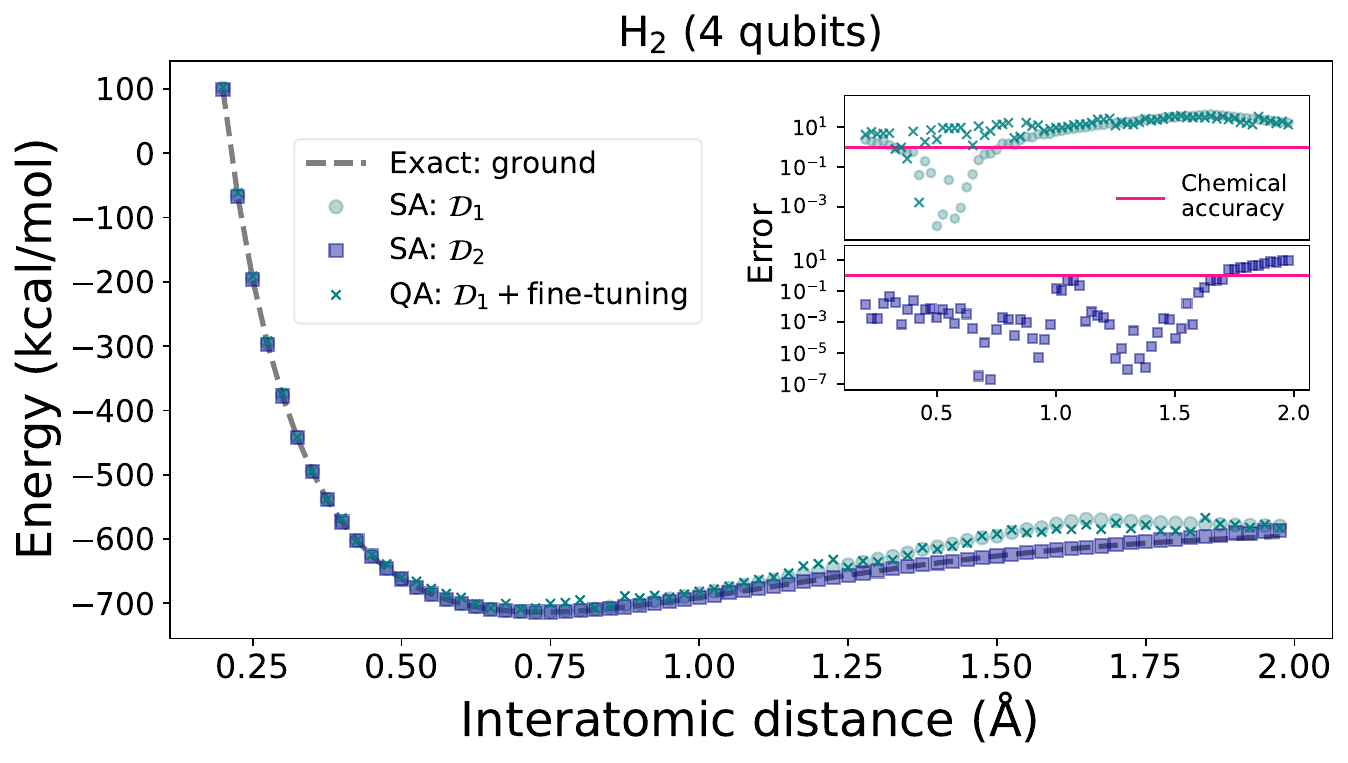}  
    \label{fig:h2_1}
    }
    \hspace{1mm}
  \subfigure[]{
    \centering
\includegraphics[width=1.43\columnwidth]{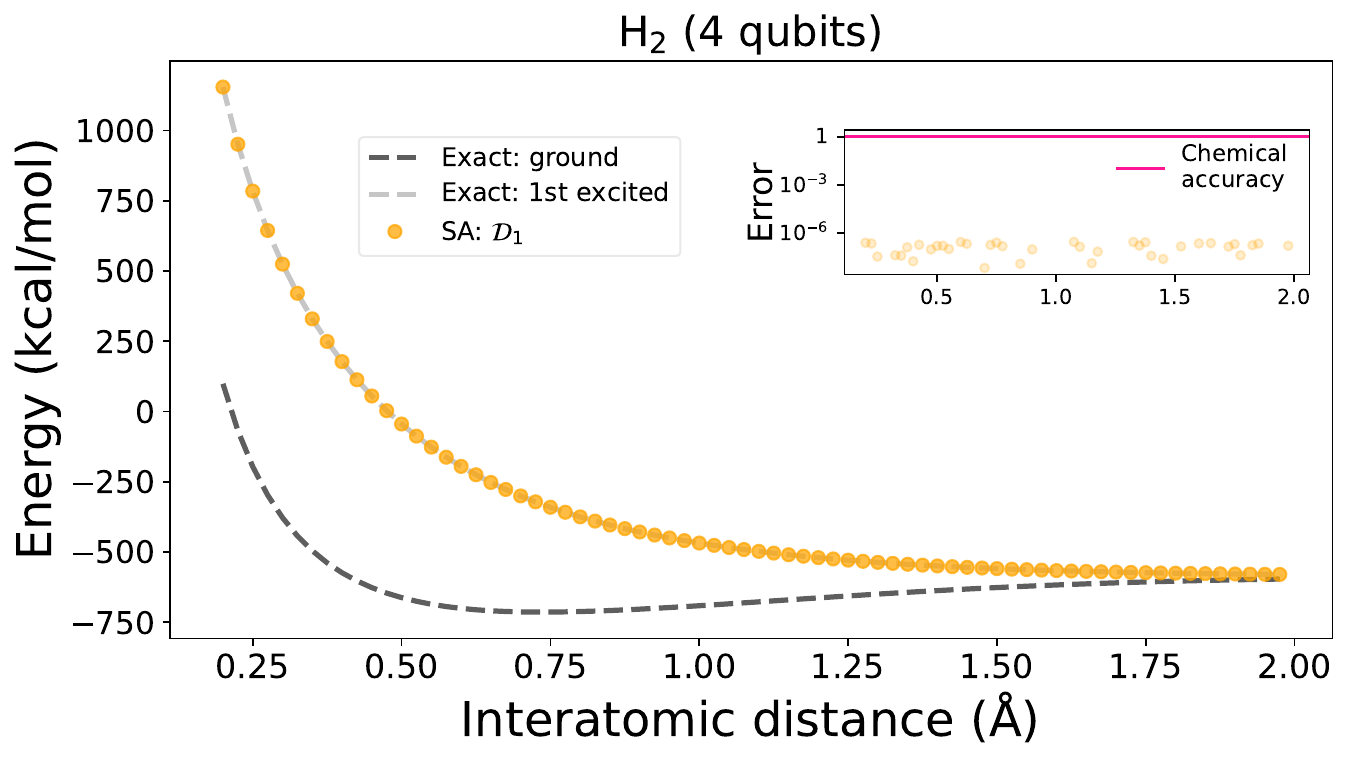} 
    \label{fig:h2_2}
    }
  \caption{Applications of the VarQA algorithm to calculations of the $\mathrm{H_2}$ ground and excited states, for a range of interatomic distances from $0.200\mathrm{\AA}$ to $1.975\mathrm{\AA}$. (a) SA results for the $\mathrm{H_2}$ ground state, with ansatz generated from digitizers $\mathcal{D}_1$ and $\mathcal{D}_2$; QA results for the $\mathrm{H_2}$ ground state, using the optimal digitizer $\mathcal{D}_1$ predicted from SA, followed by further fine-tuning. Insets display the errors relative to the exact $\mathrm{H_2}$ ground state energies. (b) SA results for the first physical excited state of $\mathrm{H_2}$, ($^3\Sigma^+_u$), with ansatz generated from digitizer $\mathcal{D}_1$.  The chemical accuracy threshold shown in the insets is $\SI{1}{\kcal\per\mole}$.
  } 
    \label{fig:h2}
\end{figure*}

In Fig.~\ref{fig:h2_1} we present the ground state energy calculations for the H$_2$ molecule using the VarQA algorithm through SA with digitizers $\mathcal{D}_1$ and $\mathcal{D}_2$, and through QA using digitizer $\mathcal{D}_1$ and limited fine-tuning, for a range of interatomic distances.
The calculated ground state energies were compared to the exact values obtained from exact diagonalization and the results demonstrate the effectiveness of the digitizer techniques and integer variational angles. The SA results achieved through digitizer $\mathcal{D}_2$ exhibit quantitative agreement with the exact ground state energies, achieving energy values within chemical accuracy for most interatomic distances. The integer variational angles can be fine-tuned to achieve higher accuracy. However, this procedure was not performed for SA. To illustrate how such fine-tuning works, we provide an example for a specific interatomic distance in Appendix~\ref{app:digital}. The optimal integer variational angles of $\mathcal{D}_2$ for every interatomic distance are listed in Table~\ref{tab:h2g} of Appendix~\ref{app:h2}. The SA results obtained using digitizer $\mathcal{D}_1$ show qualitative agreement with the exact ground state energies, except for an interatomic distance range between $1.25\si{\angstrom}$ and $1.75\si{\angstrom}$. For certain interatomic distances, $\mathcal{D}_1$ also produces results within the chemical accuracy threshold of $\SI{1}{\kcal\per\mole}$. The QA experimental results for $\mathrm{H_2}$ from using digitizer $\mathcal{D}_1$ and limited fine tuning are comparable to the SA results obtained using digitizer $\mathcal{D}_1$. This suggests that the optimal $\mathcal{D}_1$ digitizer parameters from SA serve as effective warm-start angles for QA.
The small differences between QA and SA results might indicate that the QPU sampling statistics in QA is different from the Boltzmann statistics in SA. However, our algorithm does not presuppose a Boltzmann distribution and can be applied to other distributions as well. Therefore, QA results can be further optimized with sufficient access to annealing hardware, allowing for a more thorough search for the optimal variational parameters.

\begin{figure*}[htb!]  
  \subfigure[]{
  \centering
\includegraphics[width=0.99\columnwidth]{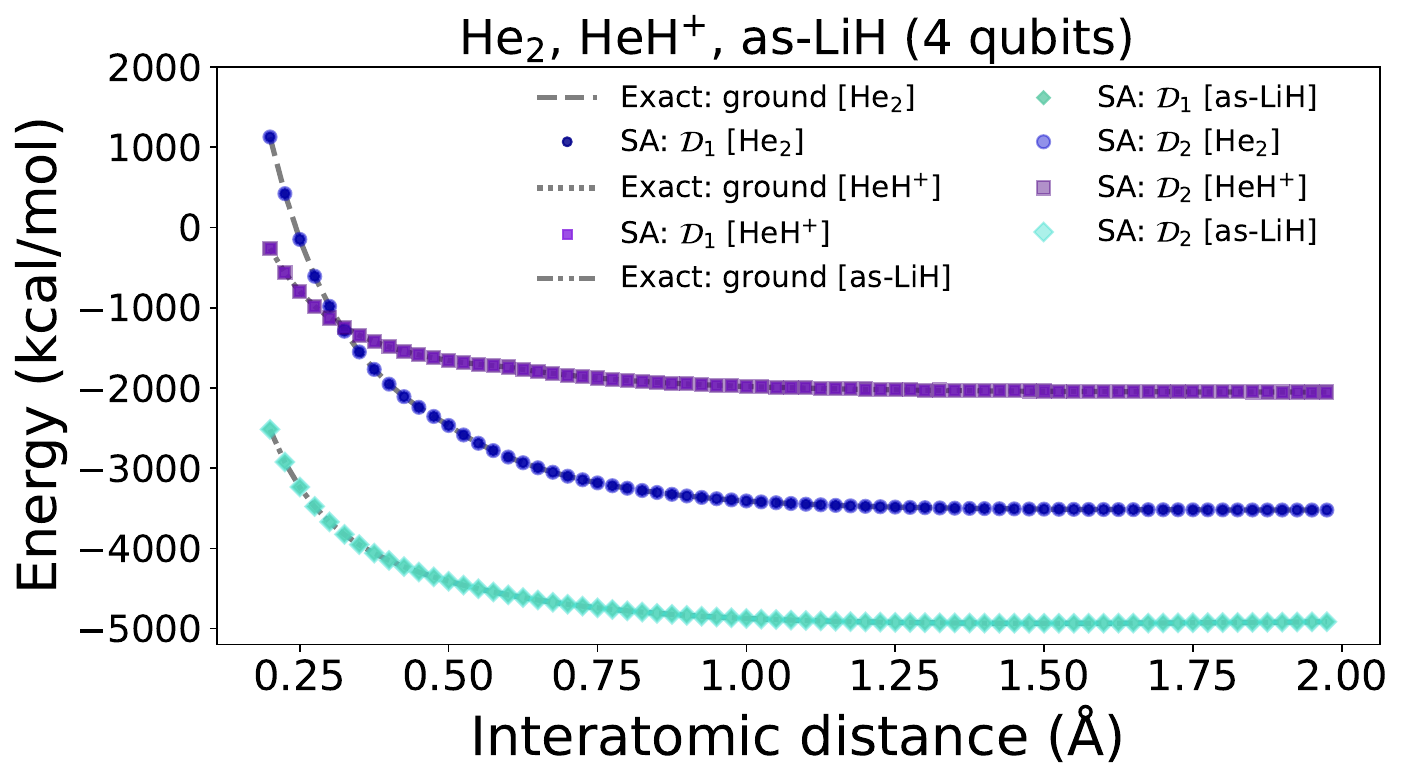}  
    \label{fig:he2_1}
    }
    \hspace{0mm}
  \subfigure[]{
    \centering
\includegraphics[width=0.99\columnwidth]{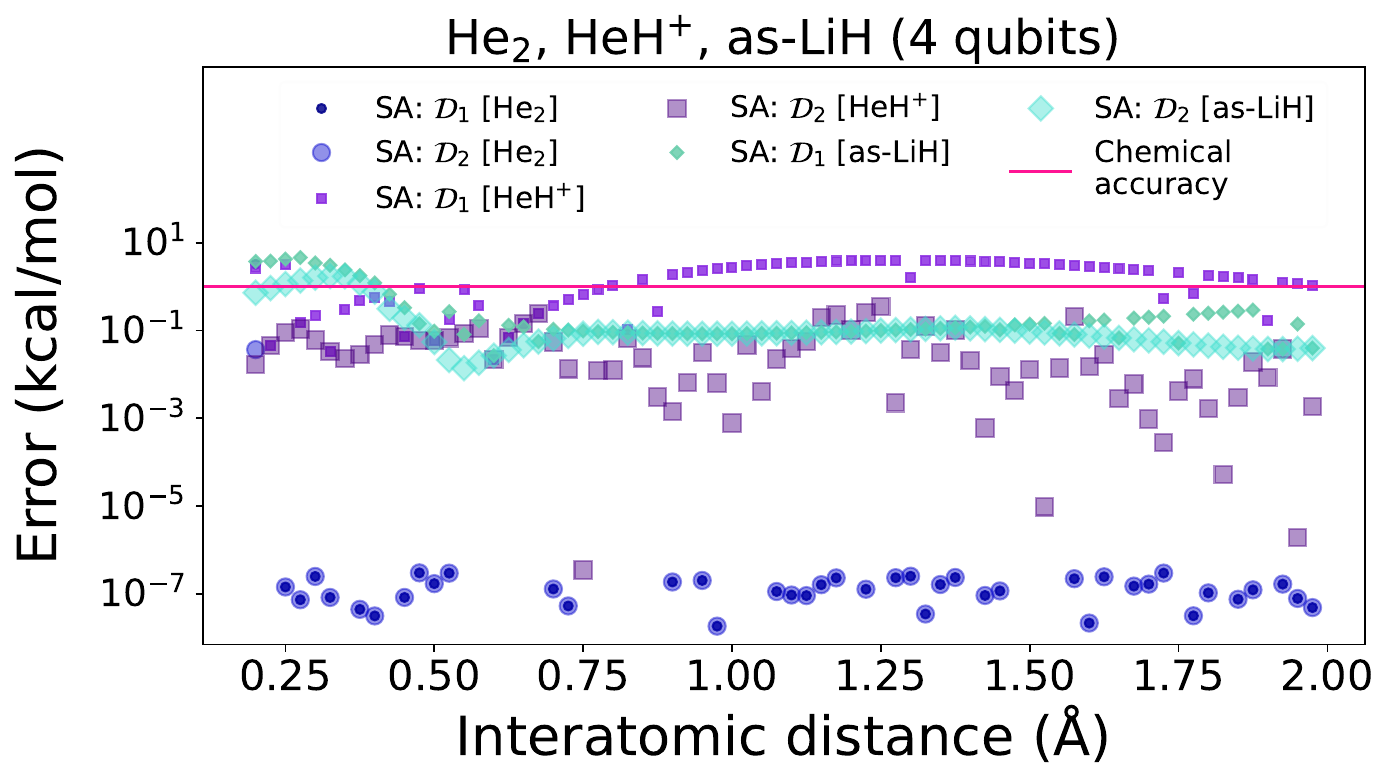} 
    \label{fig:he2_2}
    }
  \caption{Applications of the VarQA algorithm to calculations of the ground states of $\mathrm{He_2}$, $\mathrm{HeH^{+}}$, $\mathrm{as\text{-}LiH}$, for a range of interatomic distances from $0.200\mathrm{\AA}$ to $1.975\mathrm{\AA}$. (a) SA results for the ground states of $\mathrm{He_2}$, $\mathrm{HeH^{+}}$, $\mathrm{as\text{-}LiH}$, with ansatz generated from digitizers $\mathcal{D}_1$ and $\mathcal{D}_2$.   (b) The errors relative to the exact ground state energies of $\mathrm{He_2}$, $\mathrm{HeH^{+}}$, $\mathrm{as\text{-}LiH}$. The chemical accuracy threshold shown in the insets is $\SI{1}{\kcal\per\mole}$.} 
    \label{fig:he2}
\end{figure*}

\begin{figure*}[htb!]  
  \subfigure[]{
  \centering
\includegraphics[width=0.99\columnwidth]{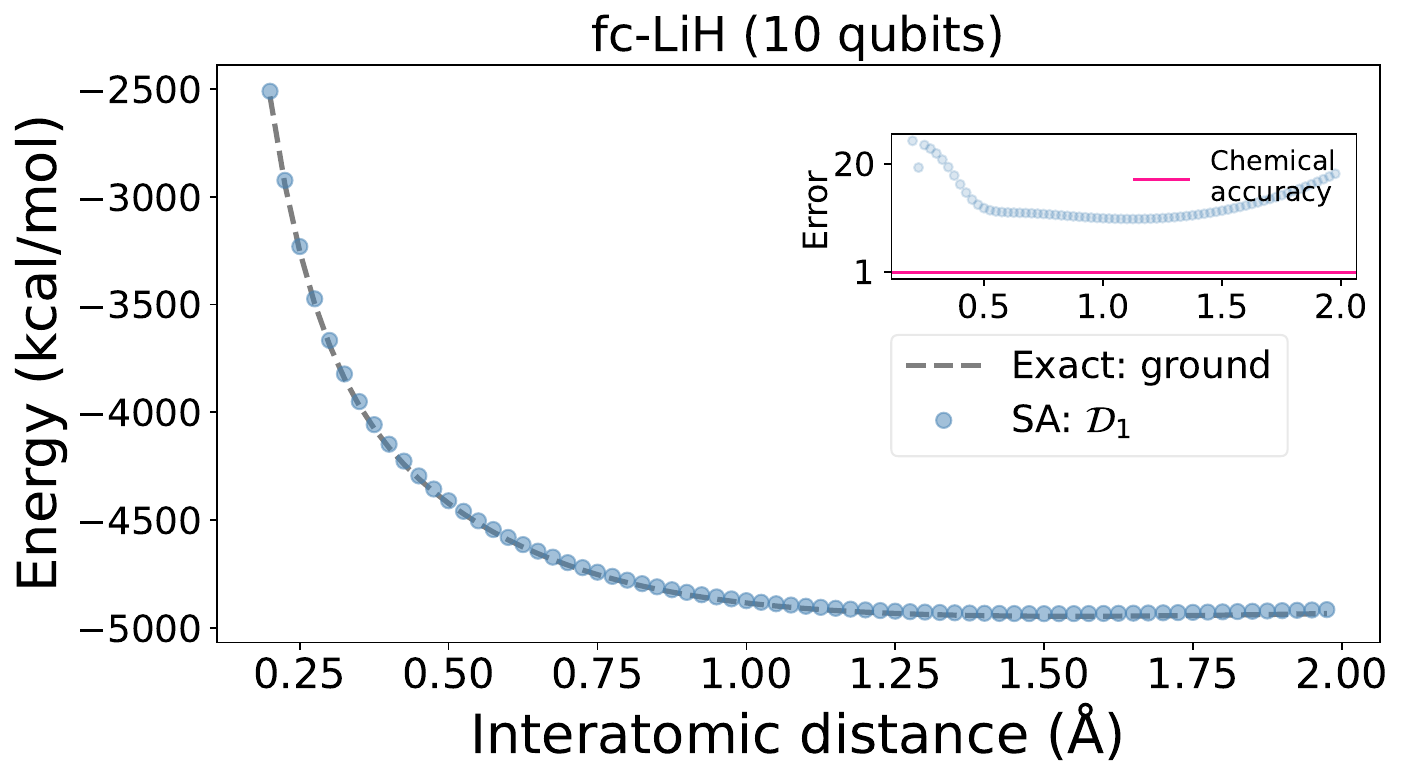}  
    \label{fig:fc_lih_1}
    }
    \hspace{0mm}
  \subfigure[]{
    \centering
\includegraphics[width=0.99\columnwidth]{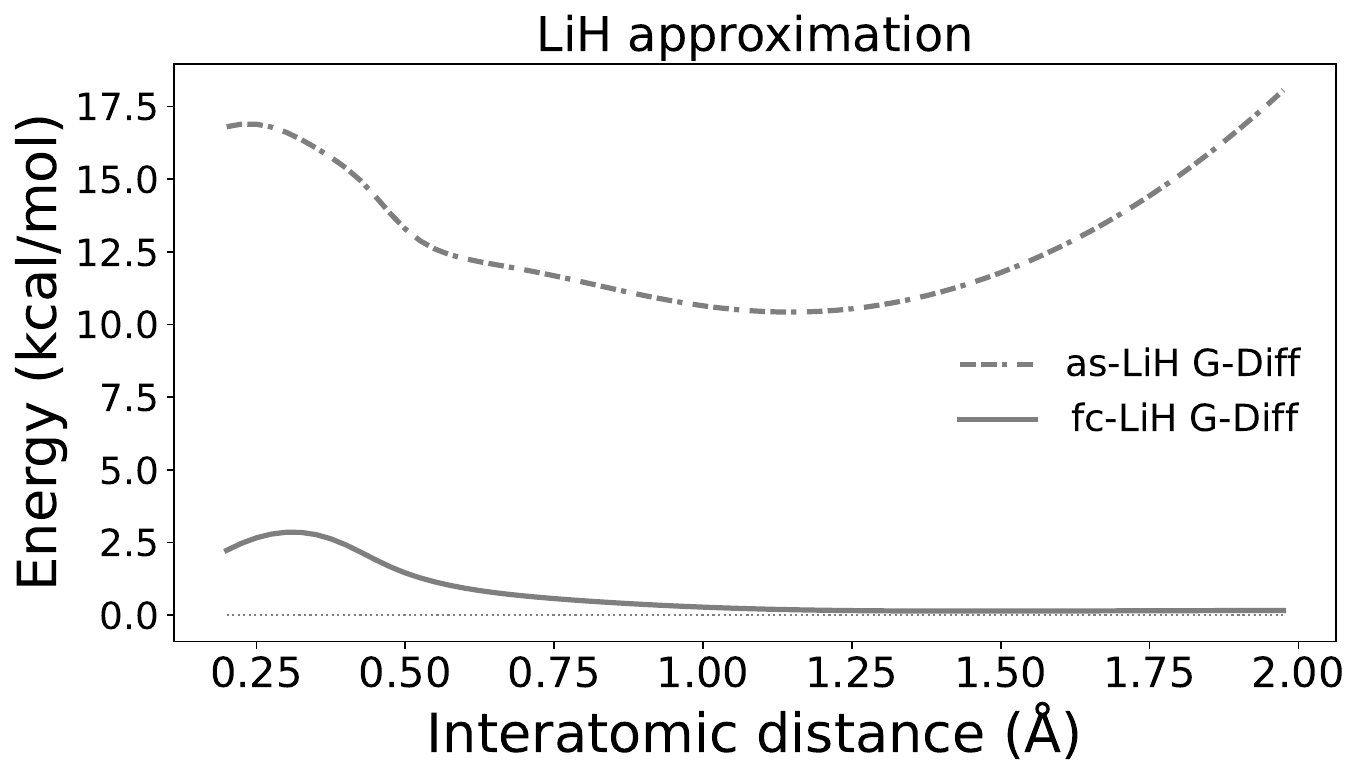} 
    \label{fig:as_lih_fc_lih_lih}
    }
  \caption{(a) SA results for the ground states of $\mathrm{fc\text{-}LiH}$, with ansatz generated from digitizers $\mathcal{D}_1$, for a range of interatomic distances from $0.200\mathrm{\AA}$ to $1.975\mathrm{\AA}$.  (b) The assumption errors from active space selection and the frozen core approximation, of the LiH molecule, are defined as the differences (G-Diff) between their (i.e., $\mathrm{as}\text{-}\mathrm{LiH}$'s and $\mathrm{fc}\text{-}\mathrm{LiH}$'s) ground state energies and the ground state energies of LiH.} 
    \label{fig:fc_lih}
\end{figure*}

Similarly, we computed the ground state energies of $\mathrm{He_2}$, $\mathrm{HeH^+}$,  $\mathrm{as}\text{-}\mathrm{LiH}$ using the VarQA algorithm through SA with digitizers $\mathcal{D}_1$ and $\mathcal{D}_2$, and the results are shown in Fig.~\ref{fig:he2_1}. The errors with respect to exact values are presented in Fig.~\ref{fig:he2_2}. As expected, digitizer $\mathcal{D}_2$ gives more accurate results than digitizer $\mathcal{D}_1$, for all three molecules. Nevertheless, most errors remain well within chemical accuracy, demonstrating the success of the VarQA algorithm and the digitizer techniques.
Note that for the errors of $\mathrm{as}\text{-}\mathrm{LiH}$, we compared the VarQA results with the exact ground state energy of $\mathrm{as}\text{-}\mathrm{LiH}$, not $\mathrm{LiH}$. If the results were to be compared with the ground state energy of $\mathrm{LiH}$, the additional errors, introduced by the active space reduction, are approximately $\SI{15}{\kcal\per\mole}$, as shown in Fig.~\ref{fig:as_lih_fc_lih_lih}. 

For the ground state energy of $\mathrm{fc}\text{-}\mathrm{LiH}$, Fig.~\ref{fig:fc_lih_1} shows that the SA results obtained using digitizer $\mathcal{D}_1$ exhibit qualitative agreement with the exact ground state energies of $\mathrm{fc}\text{-}\mathrm{LiH}$. Again, the frozen core assumption naturally introduces an approximation error compared to $\mathrm{LiH}$. As shown in Fig.~\ref{fig:as_lih_fc_lih_lih}, this assumption errors range from $\SI{0.14}{\kcal\per\mole}$ to $\SI{2.85}{\kcal\per\mole}$.

\begin{figure*}[htb!]  
  \subfigure[]{
  \centering
\includegraphics[width=0.99\columnwidth]{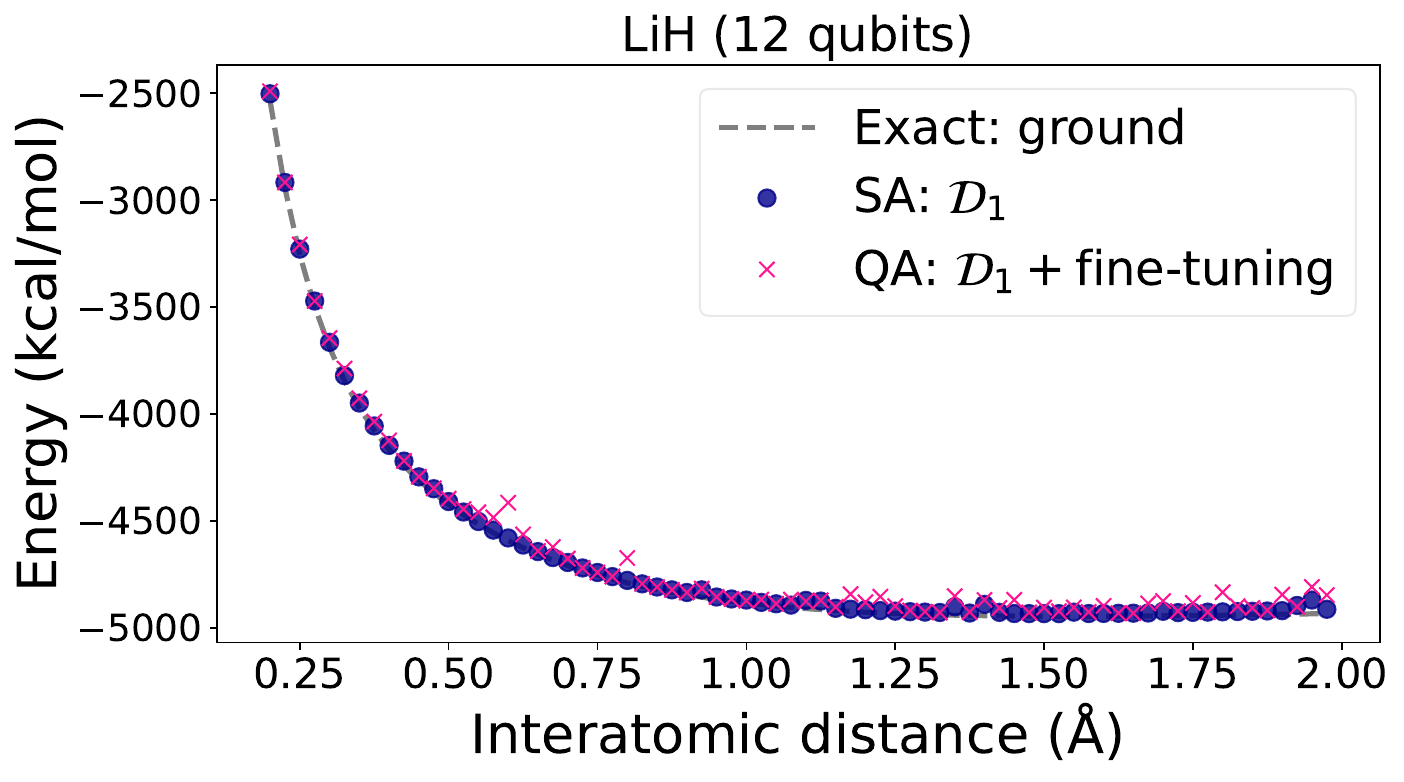}  
    \label{fig:exp_lih}
    }
    \hspace{0mm}
  \subfigure[]{
    \centering
\includegraphics[width=0.99\columnwidth]{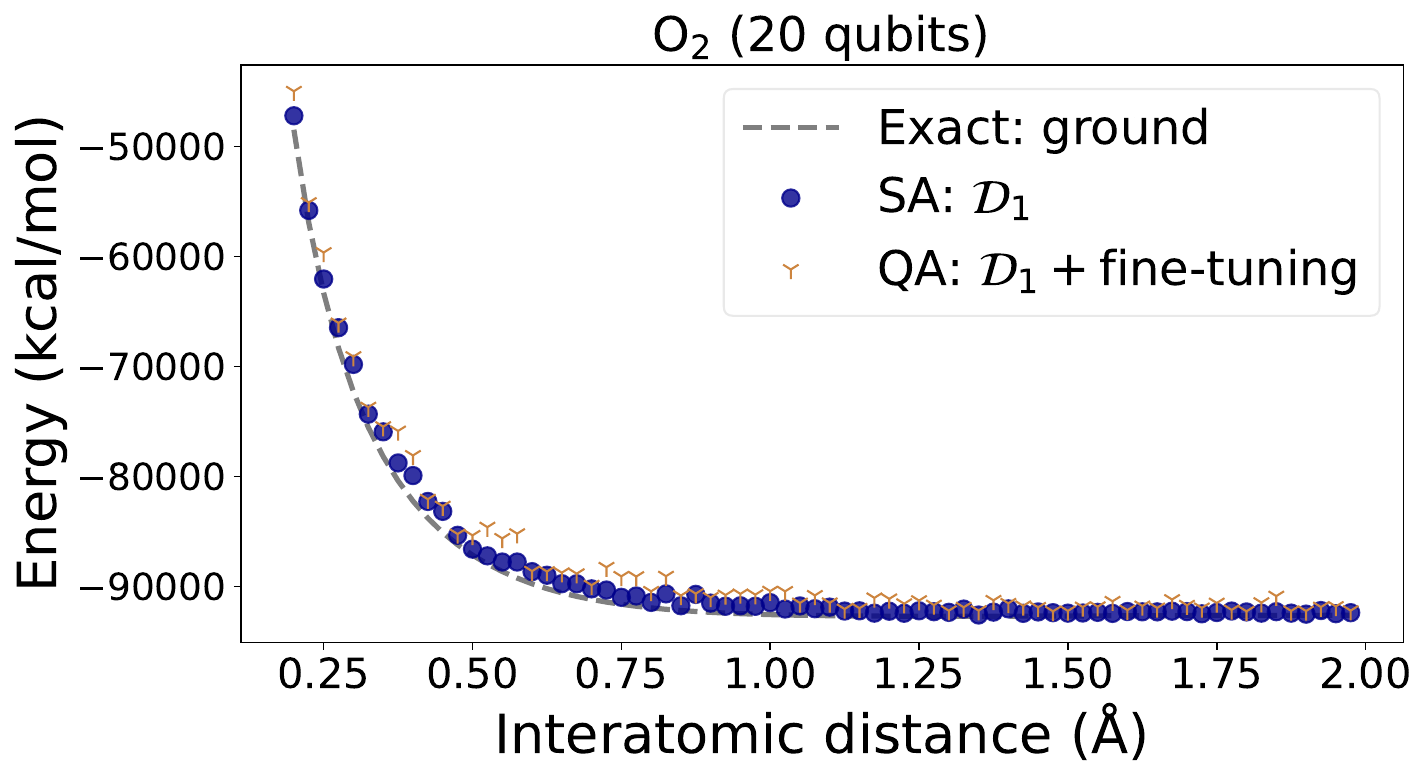} 
    \label{fig:exp_o2}
    }
  \caption{Applications of the VarQA algorithm to calculations of the ground states of $\mathrm{LiH}$ and $\mathrm{O_2}$, for a range of interatomic distances from $0.200\mathrm{\AA}$ to $1.975\mathrm{\AA}$. (a) SA results for the $\mathrm{LiH}$ ground state, with ansatz generated from digitizers $\mathcal{D}_1$; QA results for the $\mathrm{LiH}$ ground state, using the optimal digitizer $\mathcal{D}_1$ predicted from SA, followed by minimal fine-tuning to a few data points. (b) SA results for the $\mathrm{O_2}$ ground state, with ansatz generated from digitizers $\mathcal{D}_1$; QA results for the $\mathrm{O_2}$ ground state, using the optimal digitizer $\mathcal{D}_1$ predicted from SA, followed by minimal fine-tuning to a few data points.} 
    \label{fig:lih_o2}
\end{figure*}

Finally, we present the ground state energy calculations of molecules $\mathrm{LiH}$ and $\mathrm{O_2}$ using the VarQA algorithm through SA and QA with digitizer $\mathcal{D}_1$ in Fig.~\ref{fig:exp_lih} and Fig.~\ref{fig:exp_o2}, respectively. The VarQA algorithm through SA with digitizer $\mathcal{D}_1$ gives results that align with the exact solutions for these larger molecules. The optimal variational angles from SA serve as warm-start angles for QA, which yields results comparable to the values obtained through SA and the exact ground state energies after a finite amount of fine tuning. 
Due to limited QPU access privileges, most interatomic distances were evaluated using the warm-start angles directly without fine-tuning.

\subsubsection{Excited state evaluation}
Our VarQA approach can be easily extended to the electronic excited state energy calculation by using the overlap-based method~\cite{tep2019,Higgott2019variationalquantum,mcardle_quantum_2020}. To search for the $k$-th eigenstate of the Hamiltonian $H$, i.e., $\ket{e_k}$, one can define a new Hamiltonian $H^{k}$ as follows:
\begin{equation}
\label{eq:Hk}
    H^{k} = H + \sum_{i=0}^{k-1}\alpha_i\ket{e_i}\bra{e_i} \,,
\end{equation}
where $\ket{e_i}$ is the $i$-th eigenstate of $H$ with energy $e_i$.  If one chooses $\alpha_i > e_k - e_i$, the ground state of $H^{k}$ can sufficiently approximate $\ket{e_k}$.

Below we illustrate the results for the first physical excited state of $\mathrm{H_2}$. Let $H_{\mathrm{H_2}}$ be the electronic Hamiltonian of $\mathrm{H_2}$ after the Jordan-Wigner transformation (See Eq.~\ref{eq:pauliform}). Note that since $H_{\mathrm{H_2}}$ is first defined in the second-quantized language, many eigenstates are unphysical. To search for the first physical excited state of $\mathrm{H_2}$, $^3\Sigma^+_u$ (the lowest triplet state), one has to filter for states that have two electrons and zero magnetization. Which eigenstate corresponds to this state depends on the interatomic distance between the two hydrogen atoms.
Therefore, we define the new Hamiltonian $H^{k}_{\mathrm{H_2}}$ differently across three ranges of interatomic distances as follows: 
\begin{enumerate}
    \item for $d<0.475 \si{\angstrom}$, $k=5$; 
    \item for $0.475 \mathrm{\AA} \lesssim d  < 0.75 \mathrm{\AA}$, $k=3$; 
    \item for $d  \gtrsim 0.75 \mathrm{\AA}$, $k=1$. 
\end{enumerate}
We chose $\alpha_i = \SI{1255.02}{\kcal\per\mole}$ ($2$ Hartree) for all $i$
in Eq.~\ref{eq:Hk}, and applied the VarQA algorithm to the newly defined $H^{k}_{\mathrm{H_2}}$ (as $H_{\text{elec}}$ in Algorithm~\ref{alg:1}). The results are presented in 
Fig.~\ref{fig:h2_2}. With this newly defined $H^{k}_{\mathrm{H_2}}$, the VarQA algorithm can estimate the first physical excited state energy of $\mathrm{H_2}$ using the digitizer $\mathcal{D}_1$ with an error below $3\times 10^{-7}$ kcal\,mol$^{-1}$, notably within chemical accuracy. 
The optimal integer variational angles of $\mathcal{D}_1$ for every interatomic distance are listed in Table~\ref{tab:h2e} of Appendix~\ref{app:h2}.

\section{Comparison with other quantum annealing methods}
\label{sec:methodcomparison}
Other existing QA methods for quantum chemistry include the Xia-Bian-Kais (XBK) method~\cite{xia_electronic_2018} and the Quantum Annealer Eigensolver (QAE) algorithm~\cite{teplukhin2020electronic}. For both XBK and QAE, the qubit requirement increases exponentially with the problem size, which is the number of spin orbitals in this context. We recently used symmetry-adapted encodings to reduce the qubit requirement for the XBK method, but it requires the molecular system to have symmetry to be taken advantage of and it does not change the scaling~\cite{desroches2025electronicstructuretheorymolecular}. In XBK and QAE, the solution is encoded only in the lowest state of the quantum annealer, thus requiring the algorithm to execute in a closed system. The additional qubit requirements in the XBK method arise from converting $\sigma^x$ and $\sigma^y$ operators into higher-dimensional $\sigma^z$ operators, whereas for QAE, they stem from computing all elements of the full configuration interaction Hamiltonian matrix and using a fixed-point representation. Our method, VarQA, does not require an expansion over the Hilbert space of the electronic Hamiltonian. This is done by mapping the annealing process to a higher dimensional open system, and encoding the trial solution using statistics in all the higher excited states of the annealer (See Fig.~\ref{fig:schematics}).

The XBK method also requires additional quadratization to transform the problem into a QUBO formulation, introducing another layer of exponential overhead. In contrast, both QAE and VarQA do not require quadratization. A comparison of our method with other existing QA approaches for quantum chemistry is provided in Table~\ref{tab:compare}. This table demonstrates that the number of logical qubits needed for XBK and QAE scale exponentially and that of VarQA scales linearly in system size. As an example, we also provided the number of logical qubits required for studying the LiH electronic Hamiltonian in the STO-3G basis set for each method. 
Studying the LiH molecule using XBK and QAE methods requires $r\cdot536$ (where $r$ is an exponentially growing integer parameter) and 4950 logical qubits, respectively. 
However, our VarQA method requires only 12 logical qubits for studying the same molecule. For all methods, the number of physical qubits practically realized on a quantum annealer depends on the QPU’s physical qubit connectivity relative to the Hamiltonian structure and the graph embedding techniques employed.

\begin{table}
\small
\newcolumntype{?}{!{\vrule width 1.5pt}}
\begin{tabular}{|c?l|l|l|}
 \hline
 & XBK & QAE & \textsf{VarQA}\\
 \hline
 Qubits   & Exponential & Exponential & Linear\\
 \hline
 LiH & $r\cdot536$ & $4950$ &  $12$ \\
 \hline
\end{tabular}
\caption{The logical qubit scaling for XBK, QAE, and VarQA algorithms, respectively. An example of the required logical qubit resources to simulate the ground state of LiH with Hamiltonian in the STO-3G basis using these methods is given. For the XBK method, $r$ is an exponentially-growing integer parameter that controls the accuracy.}
\label{tab:compare}
\end{table}

\section{$\alpha$-VarQA}
\label{sec:alpha-varaq}
To further improve the accuracy of VarQA, multiple annealing experiments with different parameters can be conducted $\alpha$ times before constructing the trial wavefunction. To this end, we propose a multiple ($\alpha$) annealing experiment protocol, $\alpha$-VarQA, in state construction to achieve maximum accuracy. This in turn expands the parameter space from $\nu$ to $\alpha \nu$, where $\nu = M(M+1)/2+1$.

Define an ensemble of angle sets as 
\begin{equation}
\bm{\Theta}:=(\bm{\theta}^{1},\bm{\theta}^{2},\dots,\bm{\theta}^{\alpha})\,.
\end{equation}
Each set of angles has a length of $\nu$. For each set of angles, $\bm{\theta}^{k}$, its corresponding Ising Hamiltonians is:
\begin{align}
\label{eq:alpha-varaq_parameterizedHamiltonian_1}
H(\bm{\theta}^k) &= \sum_{i=1}^{M}\theta^k_{i}\sigma_{i}^{z} + \sum_{j>i=1}^{M}\theta^k_{ij}\sigma_{i}^{z}\sigma_{j}^{z} + \theta^k_0 I \,,
\end{align}
which yields the measurement probability distribution $p^k(m)$ for the bit string $m$.
For $\alpha$ sets of angles,  ensemble averaging of the measurement outcomes can be performed before constructing the parameterized trial state (Eq.~\ref{eq:trial_state}):
\begin{equation}
    \label{eq:alpha-varaq_trial_state}
\ket{\psi(\bm{\Theta})}_{\pm} = \sum_{m}\varepsilon_m\sqrt{\sum_{k=1}^{\alpha}\frac{p^{k}(m)}{\alpha}}\ket{m},\,\varepsilon_m=\pm 1\,. 
\end{equation}

Eq.~\ref{eq:alpha-varaq_trial_state} represents a coherent superposition of measurement outcomes from $\alpha$ annealing experiments, with the statistics evenly divided by $\alpha$ to maintain normalization. With ensemble averaging over different sets of angles, a greater variety of measurement outcomes is obtained, enhancing the trial state's representation of additional nonzero terms in the exact electronic state. 

Given the highly sparse nature of the second-quantized Hamiltonian, where the number of nonzero terms grows polynomially with the size of the molecule, we found that for small molecules, $\alpha=1$ is sufficient to capture these nonzero terms. This is because the size of the parameter space, $\nu$, also grows polynomially. However, to further improve accuracy, one can go beyond $\alpha=1$, particularly for large molecules.

\section{Conclusion}
\label{sec:conclusion}
In this work, we presented the VarQA algorithm, a hybrid quantum-classical approach for studying the electronic structure of molecular systems. This algorithm enables efficient solution of the electronic Hamiltonian of molecules by leveraging quantum annealing and a digitization strategy. We tested our algorithm on molecules including $\mathrm{H_2}$, $\mathrm{He_2}$, $\mathrm{HeH^+}$, $\mathrm{LiH}$, and $\mathrm{O_2}$, using the STO-3G basis set. We demonstrated that the digitizer technique ($\mathcal{D}_1$ and $\mathcal{D}_2$) provides a structured way to discretize and optimize variational parameters. We benchmarked our VarQA algorithm on classical hardware using simulated annealing method, and demonstrated the effectiveness of the VarQA algorithm on D-Wave's quantum annealer. For small molecules, such as $\mathrm{H_2}$, $\mathrm{He_2}$, and $\mathrm{HeH^+}$, the ground and first excited state energies computed using VarQA and the digitizer technique are within chemical accuracy. 

A key challenge in applying quantum annealing to quantum chemistry problems is due to %}
the quantum Ising Hamiltonian format that imposes certain restrictions, allowing only quadratic interactions between Pauli-Z operators. While this structure facilitates simpler manipulation of qubit systems, such as through ZZ couplers, and enables relatively simpler scalability to large quantum devices, it often requires additional overhead to encode problems into quantum Ising machines or quantum annealers. These qubit resource overheads can be exponential~\cite{xia_electronic_2018,teplukhin2020electronic}, making studying larger molecules impossible. 
In quantum chemistry, where the second-quantized Hamiltonian is often sparse, we circumvent these traditional encoding overheads by leveraging statistical properties of higher excited states in quantum annealers to encode the electronic structure problem of quantum chemistry. 

By taking advantage of distinct feature of the Ising energy landscape, we demonstrate the use of digitizers as a proof of concept to conduct a parallel  search for the optimizer parameters.
Our future work will focus on developing more sophisticated optimization techniques to enhance the efficiency of global search strategies. Even though the digitizer technique provides a structured and parallelizable approach for exploring the parameters space, these advanced methods could further refine variational parameters and improve energy convergence. One option is to leverage quantum annealing again for this optimization.

Our VarQA approach, combined with digitization and potentially other optimization techniques, are generally applicable to solving the eigenvalue problem of sparse Hamiltonians, making it a general strategy for a wide range of eigenvalue problems.

\section*{Acknowledgement}
This material is based upon work supported by the U.S. Department of Energy, Office of Science, Office of Advanced Scientific Computing Research, under Award Number DE-SC0024216. We thank Northeastern University Research Computing for providing classical computing resources and D-Wave, Inc. for providing quantum computing resources. We thank Jack Raymond for useful discussions. 

KYA ©2025 The MITRE Corporation. ALL RIGHTS RESERVED. Approved for public release. Distribution unlimited. PR$-$25$-$0829.

\bibliography{refs}

\onecolumngrid
\appendix
\section{Fine-tuning integer Ising coefficients}
\label{app:digital}
We consider the evaluation of $\mathrm{H_2}$ ground state using the digitizer $\mathcal{D}_2$.
For the interatomic distance of $d = 1.950\mathrm{\AA}$, the optimal integer variational angles for SA are 
$\bm{\theta^{*}} =\left\{\theta_1^{*}=0,\theta_2^{*}=0,\theta_3^{*}=1,\theta_4^{*}=0,\theta_{12}^{*}=1,\theta_{13}^{*}=-1,\theta_{14}^{*}=1,\theta_{23}^{*}=1,\theta_{24}^{*}=-1,\theta_{34}^{*}=1,\theta_{0}^{*}=1\right\}$. As shown in Fig.~\ref{fig:h2_1}, at $d = 1.950\mathrm{\AA}$, the error to the exact ground state is above the chemical accuracy threshold.

We aim to fine-tune $\theta_2$ beyond the integer value. Starting from its original optimal value $0$, one can slowly vary $\theta_2$ in both negative and positive directions. By applying the VarQA algorithm to the new angle set, a new expected energy can be obtained. The process then iterates in the direction that results in a lower expected energy, until an optimal point where the expected energy cannot be lowered. Such fine-tuning can be guided using optimization technique like gradient descent.

\begin{figure*}[htb!]  
  \subfigure[]{
\includegraphics[width=0.48\columnwidth]{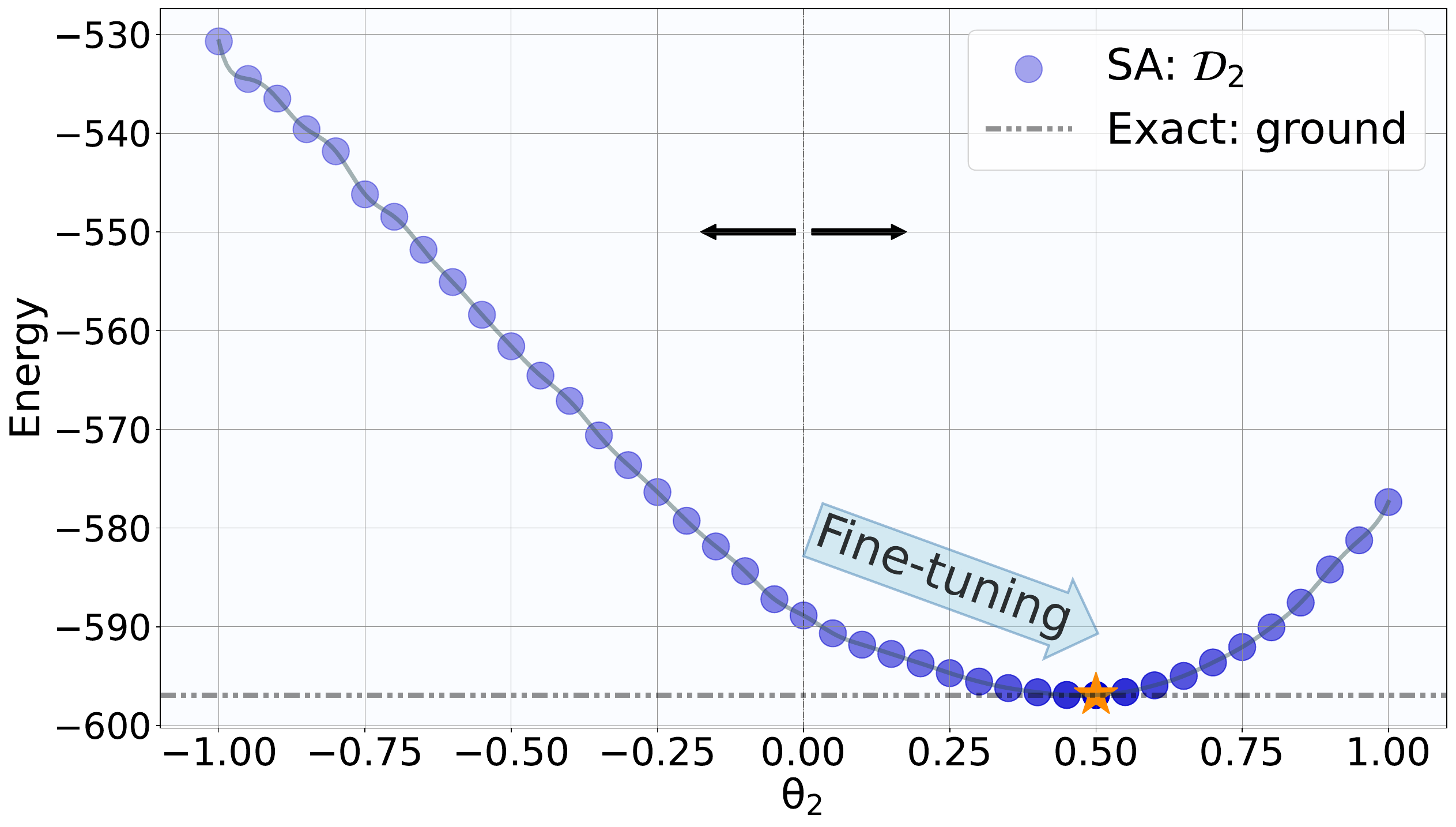}  
    }
    \hspace{0mm}
  \subfigure[]{
\includegraphics[width=0.48\columnwidth]{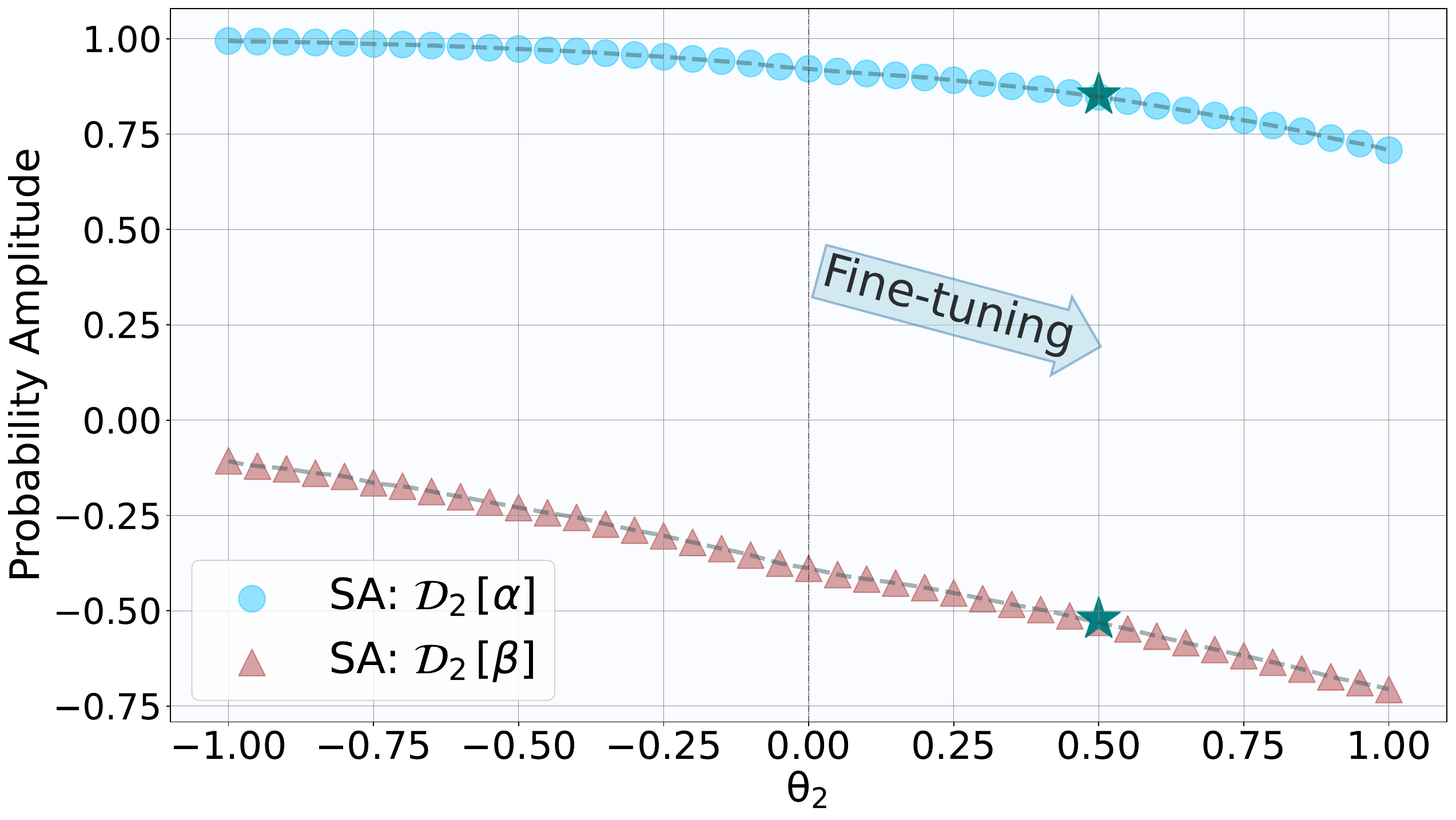} 
    }
  \caption{Starting from the optimal integer angle $\theta_2=0$, fine-tuning it in the positive direction and iterating the VarQA algorithm, until $\theta_2=0.5$, leads to: (a) a lower expected energy, and  (b) the coefficients $\alpha$ and $\beta$ of the trial wavefunction aligning with those of the exact $\mathrm{H_2}$ ground-state wavefunction at an interatomic distance of $1.950\mathrm{\AA}$.}
  \label{fig:digital}
\end{figure*}

In big-endian ordering, the exact ground state of $\mathrm{H_2}$ can be expressed as:
\begin{equation}
    \label{eq:h2gs}
\ket{\psi_g}_{\mathrm{H_2}} = \alpha|0101\rangle +  \beta|1010\rangle \,,
\end{equation}
where the values of $\alpha$ and $\beta$ depend on the interatomic distance. At the interatomic distance of $1.950\mathrm{\AA}$, $\alpha \approx 0.8529$ and $\beta \approx -0.5220$. We illustrate the conceptual procedure in Fig.~\ref{fig:digital}. We also label the exact ground state energies and the exact probability amplitudes of $\mathrm{H_2}$, at the interatomic distance of $1.950\mathrm{\AA}$.

After fine-tuning a single angle ($\theta_2$), the error decreases from $\SI{8.0877}{\kcal\per\mole}$ to $\SI{0.0342}{\kcal\per\mole}$, falling
well below the threshold for chemical accuracy.

\section{Relating \texorpdfstring{$\langle\psi(\bm{\theta})|H_{\text{elec}}|\psi(\bm{\theta})\rangle$}{expectation value} 
\label{app:exp_eval}
to annealing computational basis measurement}
Recall that the trial state (Eq.~\ref{eq:trial_state}) is
\begin{equation}
    \label{eq:trial_state_app}
    \ket{\psi(\bm{\theta})}_{\pm} = \sum_{m}\varepsilon_m \sqrt{p(m)}\ket{m},\,\,\text{where  }\varepsilon_m=\pm 1\,.
\end{equation}
We estimate each $p(m)$ from a sample of measurements in the computational basis at the end of anneal:
\begin{equation}   
\label{eq:trial_state_exp_app}
    \ket{\psi(\bm{\theta})}_{\pm} = \sum_{m}\varepsilon_m \sqrt{\frac{\sum_{s=1}^{S}[m^s=m]}{S}}\ket{m},\,\varepsilon_m=\pm 1\,.
\end{equation}
Therefore, we are effectively performing the computational basis sampling described in~\cite{PhysRevResearch.4.033173}, which was shown to be effective for estimating quantum expectation values.
We can relate the quantum expectation-value ${}_{\pm}\langle\psi(\bm{\theta})|H_{\text{elec}}|\psi(\bm{\theta})\rangle_{\pm}$ to the probability elements $p(m)$ as:

\begin{align}
{}_{\pm}\langle\psi(\bm{\theta})|H_{\text{elec}}|\psi(\bm{\theta})\rangle_{\pm} 
&= {\sum_{m,n}}^{\prime}|\langle m|\psi(\bm{\theta})\rangle|^2|\langle n|\psi(\bm{\theta})\rangle|^2 \frac{\langle m |H_{\text{elec}}|n \rangle}{\langle m|\psi(\bm{\theta})\rangle\langle \psi(\bm{\theta})|n\rangle} \nonumber\\
&= {\sum_{m,n}}^{\prime}\varepsilon_m \varepsilon_n \sqrt{p(m)} \sqrt{p(n)}\langle m |H_{\text{elec}}|n \rangle \,,
    \label{eq:exp_eval_3}
\end{align}
where the primed summation~\cite{PhysRevResearch.4.033173} indicates the exclusion of terms 
with $p(m) = 0$ or $p(n) = 0$.

The electronic Hamiltonian $H_{\text{elec}}$ consists of a sum of Pauli strings, as given by Eq.~\ref{eq:pauliform}:
\begin{equation}
\label{eq:pauliform_app}
    H_{\text{elec}} = \sum_{i}\gamma_iP_i \,.
\end{equation}
The corresponding transition element can be expressed as:
\begin{align}
\langle m |H_{\text{elec}}|n \rangle
&= \sum_{i}\gamma_i\langle m |P_i|n \rangle\,.
\label{eq:exp_eval_3_app}
\end{align}
As shown in~\cite{PhysRevResearch.4.033173}, with $P_i$ being products of Pauli operators acting on individual qubits and $\ket{m},\ket{n}$ being computational basis states, the matrix elements $\langle m|P_i|n \rangle$ can be efficiently evaluated using algebraic methods. The required classical computational time scales linearly with the number of qubits, which is the number of orbitals $M$ in our case. Therefore, $\langle m|H_{\text{elec}}|n \rangle$ can be efficiently computed using classical methods, as $H_{\text{elec}}$ is typically sparse and consists of a polynomial number of Pauli strings $P_i$ in relation to the number of qubits. As discussed in Sec.~\ref{sec:parameterized_construction}, the number of nonzero terms in the trial state scales polynomially with $M$. Consequently, the number of terms in the primed summation of Eq.~\ref{eq:exp_eval_3_app} is also polynomial in $M$. 

\subsection{Relating \texorpdfstring{$\langle\psi(\bm{\theta})|H_{\text{elec}}|\psi(\bm{\theta})\rangle$}{expectation value} to the final state of anneal \texorpdfstring{$\rho(t_f,\theta)$}{rho}}
Recall that for the computational or Ising basis measurement, the probability element $p(m)$ is related to the final state of anneal $\rho(t_f,\bm{\theta})$ as follows:
\begin{equation}
    \label{eq:p_2_app}
    p(m)= \langle m|\rho(t_f,\bm{\theta})|m\rangle\,.
\end{equation}
Therefore, we can establish the following relationship between the quantum expectation value of electronic Hamiltonian energy and the final state of annealing.
\begin{equation}
    {}_{\pm}\langle\psi(\bm{\theta})|H_{\text{elec}}|\psi(\bm{\theta})\rangle_{\pm} = {\sum_{m,n}}^{\prime}\varepsilon_m \varepsilon_n \sqrt{\langle m|\rho(t_f,\bm{\theta})|m\rangle} \sqrt{\langle n|\rho(t_f,\bm{\theta})|n\rangle}\langle m |H_{\text{elec}}|n \rangle \,,
\end{equation}
where, once again, the primed summation denotes the exclusion of vanishing diagonal elements in $\rho(t_f,\bm{\theta})$.

\section{Effect of the number of trials $T$}
\label{app:trials_samples}
\begin{figure}[htb!]
    \centering
\includegraphics[width=0.78\columnwidth]{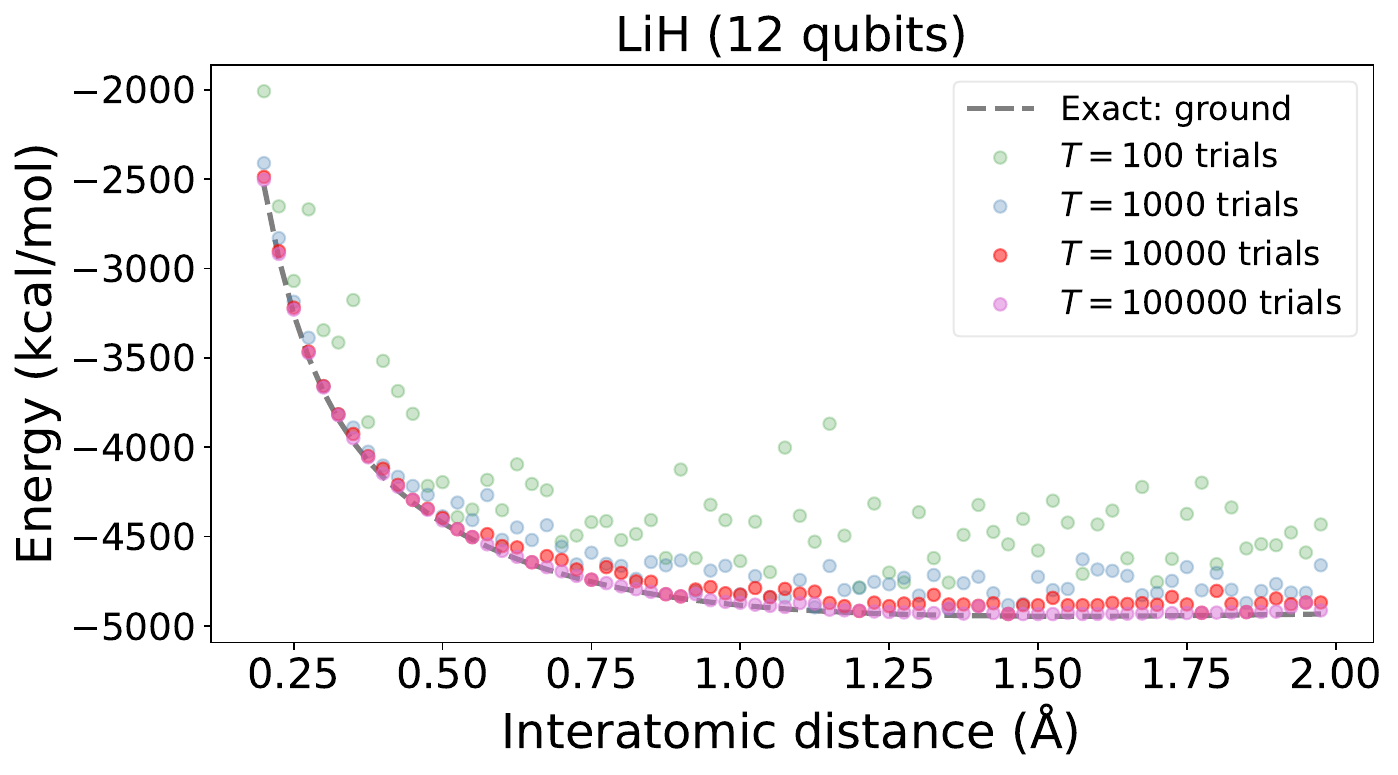}
    \caption{The number of samples $S=1000$. As the number of trials $T \in \{100,1000,10000,100000\}$ of random angles of $\mathcal{D}_1$ increases, the expected energy approaches the exact ground state energy of LiH.}
    \label{fig:T}
\end{figure}

We present data demonstrating how the number of trials, $T$, impacts the results of the VarQA algorithm, using LiH with digitizer $\mathcal{D}_1$ as an example. In each trial, a set of angles is randomly drawn from the digitizer. Fig.~\ref{fig:T} shows that as the number of trials increases, the expected energy is likely to approach the exact ground state energy.

\section{Optimal Ising parameters for $\mathrm{H_2}$ ground and excited states}
\label{app:h2}
\begin{table}[htb!]
\centering
\caption{Optimal $\mathcal{D}_2$ parameters for $\mathrm{H_2}$ ground state evaluation using SA, for every interatomic distance.}
\setlength\extrarowheight{-2pt}
\begin{tabular}{cccccccccccc}
\hline
$\theta_1$ & $\theta_2$ & $\theta_3$ & $\theta_4$ & $\theta_{12}$ & $\theta_{13}$ & $\theta_{14}$ & $\theta_{23}$ & $\theta_{24}$ & $\theta_{34}$ & $\theta_0$ & Interatomic distance ($\mathrm{\AA}$)\\
 $0$ & $-1$ & $0$ & $-1$ & $1$ & $-1$ & $0$ & $1$ & $-1$ & $1$ & $1$ & $0.200$ \\
 $0$ & $-1$ & $0$ & $-1$ & $1$ & $-1$ & $1$ & $0$ & $0$ & $1$ & $-1$ & $0.225$ \\
 $-1$ & $-1$ & $1$ & $-1$ & $1$ & $-1$ & $1$ & $0$ & $0$ & $1$ & $-1$ & $0.250$ \\
 $0$ & $-1$ & $0$ & $-1$ & $0$ & $-1$ & $1$ & $1$ & $0$ & $1$ & $0$ & $0.275$ \\
 $-1$ & $-1$ & $1$ & $-1$ & $1$ & $-1$ & $1$ & $1$ & $-1$ & $1$ & $1$ & $0.300$ \\
 $0$ & $-1$ & $0$ & $-1$ & $1$ & $-1$ & $1$ & $1$ & $0$ & $1$ & $0$ & $0.325$ \\
 $0$ & $-1$ & $0$ & $-1$ & $1$ & $-1$ & $1$ & $1$ & $0$ & $1$ & $-1$ & $0.350$ \\
 $0$ & $-1$ & $0$ & $-1$ & $1$ & $-1$ & $1$ & $1$ & $0$ & $0$ & $-1$ & $0.375$ \\
 $0$ & $-1$ & $0$ & $-1$ & $1$ & $-1$ & $1$ & $1$ & $0$ & $1$ & $1$ & $0.400$ \\
 $-1$ & $-1$ & $1$ & $-1$ & $1$ & $-1$ & $1$ & $1$ & $-1$ & $1$ & $-1$ & $0.425$ \\
 $0$ & $-1$ & $0$ & $-1$ & $1$ & $-1$ & $1$ & $1$ & $-1$ & $1$ & $0$ & $0.450$ \\
 $-1$ & $-1$ & $1$ & $-1$ & $1$ & $-1$ & $1$ & $1$ & $-1$ & $1$ & $0$ & $0.475$ \\
 $0$ & $-1$ & $1$ & $0$ & $1$ & $-1$ & $0$ & $1$ & $-1$ & $1$ & $1$ & $0.500$ \\
 $-1$ & $-1$ & $1$ & $-1$ & $1$ & $-1$ & $1$ & $1$ & $-1$ & $1$ & $1$ & $0.525$ \\
 $-1$ & $-1$ & $1$ & $-1$ & $1$ & $-1$ & $1$ & $1$ & $-1$ & $1$ & $0$ & $0.550$ \\
 $0$ & $-1$ & $1$ & $0$ & $1$ & $-1$ & $0$ & $1$ & $-1$ & $1$ & $1$ & $0.575$ \\
 $0$ & $-1$ & $-1$ & $-1$ & $1$ & $0$ & $1$ & $1$ & $0$ & $1$ & $0$ & $0.600$ \\
 $0$ & $-1$ & $0$ & $-1$ & $1$ & $-1$ & $1$ & $1$ & $-1$ & $1$ & $-1$ & $0.625$ \\
 $-1$ & $-1$ & $0$ & $-1$ & $1$ & $0$ & $1$ & $1$ & $0$ & $1$ & $-1$ & $0.650$ \\
 $-1$ & $-1$ & $1$ & $-1$ & $1$ & $-1$ & $1$ & $1$ & $-1$ & $1$ & $-1$ & $0.675$ \\
 $-1$ & $-1$ & $1$ & $0$ & $1$ & $-1$ & $1$ & $1$ & $0$ & $1$ & $-1$ & $0.700$ \\
 $-1$ & $-1$ & $1$ & $-1$ & $1$ & $-1$ & $1$ & $1$ & $-1$ & $1$ & $-1$ & $0.725$ \\
 $-1$ & $-1$ & $0$ & $-1$ & $0$ & $-1$ & $1$ & $1$ & $-1$ & $1$ & $-1$ & $0.750$ \\
 $-1$ & $-1$ & $1$ & $0$ & $0$ & $-1$ & $1$ & $1$ & $-1$ & $1$ & $1$ & $0.775$ \\
 $-1$ & $-1$ & $0$ & $-1$ & $1$ & $-1$ & $1$ & $0$ & $-1$ & $1$ & $-1$ & $0.800$ \\
 $-1$ & $-1$ & $0$ & $-1$ & $0$ & $-1$ & $1$ & $1$ & $-1$ & $1$ & $0$ & $0.825$ \\
 $-1$ & $-1$ & $0$ & $-1$ & $1$ & $-1$ & $0$ & $1$ & $-1$ & $1$ & $-1$ & $0.850$ \\
 $-1$ & $-1$ & $1$ & $0$ & $1$ & $-1$ & $1$ & $1$ & $-1$ & $0$ & $1$ & $0.875$ \\
 $-1$ & $0$ & $1$ & $-1$ & $1$ & $-1$ & $1$ & $0$ & $-1$ & $1$ & $0$ & $0.900$ \\
 $1$ & $1$ & $0$ & $-1$ & $1$ & $-1$ & $1$ & $1$ & $-1$ & $0$ & $1$ & $0.925$ \\
 $1$ & $0$ & $1$ & $1$ & $0$ & $-1$ & $1$ & $1$ & $-1$ & $1$ & $1$ & $0.950$ \\
 $1$ & $1$ & $0$ & $-1$ & $1$ & $0$ & $1$ & $1$ & $-1$ & $1$ & $-1$ & $0.975$ \\
 $0$ & $1$ & $1$ & $-1$ & $1$ & $0$ & $1$ & $1$ & $-1$ & $1$ & $0$ & $1.000$ \\
 $-1$ & $-1$ & $1$ & $0$ & $1$ & $-1$ & $1$ & $1$ & $-1$ & $0$ & $-1$ & $1.025$ \\
 $0$ & $-1$ & $1$ & $1$ & $0$ & $-1$ & $1$ & $1$ & $-1$ & $1$ & $-1$ & $1.050$ \\
 $1$ & $0$ & $0$ & $0$ & $1$ & $-1$ & $1$ & $0$ & $-1$ & $1$ & $1$ & $1.075$ \\
 $0$ & $0$ & $1$ & $0$ & $0$ & $-1$ & $1$ & $1$ & $-1$ & $1$ & $1$ & $1.100$ \\
 $0$ & $1$ & $1$ & $-1$ & $0$ & $-1$ & $1$ & $1$ & $-1$ & $1$ & $1$ & $1.125$ \\
 $1$ & $0$ & $0$ & $0$ & $1$ & $-1$ & $1$ & $1$ & $-1$ & $0$ & $-1$ & $1.150$ \\
 $0$ & $-1$ & $0$ & $0$ & $1$ & $-1$ & $0$ & $1$ & $-1$ & $1$ & $0$ & $1.175$ \\
 $0$ & $-1$ & $0$ & $0$ & $1$ & $-1$ & $1$ & $1$ & $-1$ & $0$ & $-1$ & $1.200$ \\
 $0$ & $-1$ & $0$ & $0$ & $1$ & $-1$ & $0$ & $1$ & $-1$ & $1$ & $-1$ & $1.225$ \\
 $0$ & $-1$ & $-1$ & $-1$ & $1$ & $-1$ & $1$ & $1$ & $-1$ & $1$ & $0$ & $1.250$ \\
 $1$ & $0$ & $-1$ & $-1$ & $1$ & $-1$ & $1$ & $1$ & $-1$ & $1$ & $1$ & $1.275$ \\
 $0$ & $-1$ & $0$ & $0$ & $1$ & $-1$ & $1$ & $0$ & $-1$ & $1$ & $0$ & $1.300$ \\
 $-1$ & $-1$ & $0$ & $-1$ & $1$ & $-1$ & $1$ & $1$ & $0$ & $1$ & $-1$ & $1.325$ \\
 $0$ & $0$ & $1$ & $0$ & $1$ & $-1$ & $1$ & $1$ & $0$ & $1$ & $0$ & $1.350$ \\
 $-1$ & $-1$ & $1$ & $0$ & $1$ & $-1$ & $1$ & $1$ & $-1$ & $1$ & $1$ & $1.375$ \\
 $-1$ & $-1$ & $1$ & $0$ & $1$ & $-1$ & $1$ & $1$ & $-1$ & $1$ & $-1$ & $1.400$ \\
 $1$ & $0$ & $-1$ & $-1$ & $1$ & $-1$ & $1$ & $1$ & $-1$ & $1$ & $0$ & $1.425$ \\
 $-1$ & $0$ & $1$ & $-1$ & $1$ & $-1$ & $1$ & $1$ & $-1$ & $1$ & $1$ & $1.450$ \\
 $1$ & $-1$ & $0$ & $1$ & $1$ & $-1$ & $1$ & $1$ & $-1$ & $1$ & $-1$ & $1.475$ \\
 $0$ & $0$ & $1$ & $0$ & $1$ & $-1$ & $1$ & $1$ & $-1$ & $1$ & $-1$ & $1.500$ \\
 $1$ & $0$ & $0$ & $0$ & $1$ & $-1$ & $1$ & $1$ & $-1$ & $1$ & $0$ & $1.525$ \\
 $1$ & $0$ & $0$ & $0$ & $1$ & $-1$ & $1$ & $1$ & $-1$ & $1$ & $-1$ & $1.550$ \\
 $0$ & $-1$ & $0$ & $0$ & $1$ & $-1$ & $1$ & $1$ & $-1$ & $1$ & $-1$ & $1.575$ \\
 $0$ & $0$ & $1$ & $0$ & $1$ & $-1$ & $1$ & $1$ & $-1$ & $1$ & $1$ & $1.600$ \\
 $0$ & $0$ & $0$ & $-1$ & $1$ & $-1$ & $1$ & $1$ & $-1$ & $1$ & $0$ & $1.625$ \\
 $0$ & $0$ & $0$ & $-1$ & $1$ & $-1$ & $1$ & $1$ & $-1$ & $1$ & $1$ & $1.650$ \\
 $0$ & $0$ & $1$ & $0$ & $1$ & $-1$ & $1$ & $1$ & $-1$ & $1$ & $1$ & $1.675$ \\
 $0$ & $0$ & $1$ & $0$ & $1$ & $-1$ & $1$ & $1$ & $-1$ & $1$ & $0$ & $1.700$ \\
 $0$ & $0$ & $1$ & $0$ & $1$ & $-1$ & $1$ & $1$ & $-1$ & $1$ & $-1$ & $1.725$ \\
 $0$ & $0$ & $0$ & $-1$ & $1$ & $-1$ & $1$ & $1$ & $-1$ & $1$ & $1$ & $1.750$ \\
 $0$ & $0$ & $0$ & $-1$ & $1$ & $-1$ & $1$ & $1$ & $-1$ & $1$ & $1$ & $1.775$ \\
 $1$ & $0$ & $0$ & $0$ & $1$ & $-1$ & $1$ & $1$ & $-1$ & $1$ & $1$ & $1.800$ \\
 $0$ & $0$ & $0$ & $-1$ & $1$ & $-1$ & $1$ & $1$ & $-1$ & $1$ & $1$ & $1.825$ \\
 $0$ & $0$ & $1$ & $0$ & $1$ & $-1$ & $1$ & $1$ & $-1$ & $1$ & $0$ & $1.850$ \\
 $0$ & $-1$ & $0$ & $0$ & $1$ & $-1$ & $1$ & $1$ & $-1$ & $1$ & $0$ & $1.875$ \\
 $0$ & $-1$ & $0$ & $0$ & $1$ & $-1$ & $1$ & $1$ & $-1$ & $1$ & $-1$ & $1.900$ \\
 $1$ & $0$ & $0$ & $0$ & $1$ & $-1$ & $1$ & $1$ & $-1$ & $1$ & $0$ & $1.925$ \\
 $0$ & $0$ & $1$ & $0$ & $1$ & $-1$ & $1$ & $1$ & $-1$ & $1$ & $1$ & $1.950$ \\
 $-1$ & $-1$ & $1$ & $1$ & $1$ & $-1$ & $0$ & $1$ & $-1$ & $1$ & $1$ & $1.975$ \\
\hline
\end{tabular}
\label{tab:h2g}
\end{table}

\begin{table}[htb!]
\centering
\caption{Optimal $\mathcal{D}_1$ parameters for $\mathrm{H_2}$ excited state evaluation using SA, for every interatomic distance.}
\setlength\extrarowheight{-2pt}
\begin{tabular}{cccccccccccc}
\hline
$\theta_1$ & $\theta_2$ & $\theta_3$ & $\theta_4$ & $\theta_{12}$ & $\theta_{13}$ & $\theta_{14}$ & $\theta_{23}$ & $\theta_{24}$ & $\theta_{34}$ & $\theta_0$ & Interatomic distance ($\mathrm{\AA}$)\\
 $-1$ & $-1$ & $1$ & $1$ & $-1$ & $1$ & $1$ & $1$ & $1$ & $-1$ & $1$ & $0.200$ \\
 $-1$ & $1$ & $1$ & $1$ & $-1$ & $1$ & $1$ & $1$ & $1$ & $-1$ & $-1$ & $0.225$ \\
 $1$ & $-1$ & $1$ & $-1$ & $-1$ & $1$ & $1$ & $1$ & $1$ & $-1$ & $-1$ & $0.250$ \\
 $1$ & $-1$ & $-1$ & $-1$ & $-1$ & $1$ & $1$ & $1$ & $1$ & $-1$ & $-1$ & $0.275$ \\
 $-1$ & $-1$ & $-1$ & $1$ & $-1$ & $1$ & $1$ & $1$ & $1$ & $-1$ & $1$ & $0.300$ \\
 $-1$ & $1$ & $-1$ & $-1$ & $-1$ & $1$ & $1$ & $1$ & $1$ & $-1$ & $-1$ & $0.325$ \\
 $-1$ & $-1$ & $1$ & $1$ & $-1$ & $1$ & $1$ & $1$ & $1$ & $-1$ & $-1$ & $0.350$ \\
 $-1$ & $1$ & $-1$ & $1$ & $-1$ & $1$ & $1$ & $1$ & $1$ & $-1$ & $1$ & $0.375$ \\
 $1$ & $-1$ & $-1$ & $-1$ & $-1$ & $1$ & $1$ & $1$ & $1$ & $-1$ & $-1$ & $0.400$ \\
 $-1$ & $-1$ & $1$ & $1$ & $1$ & $1$ & $-1$ & $-1$ & $1$ & $1$ & $1$ & $0.425$ \\
 $-1$ & $1$ & $-1$ & $-1$ & $-1$ & $1$ & $1$ & $1$ & $1$ & $-1$ & $1$ & $0.450$ \\
 $1$ & $-1$ & $1$ & $1$ & $-1$ & $1$ & $1$ & $1$ & $1$ & $-1$ & $-1$ & $0.475$ \\
 $-1$ & $-1$ & $-1$ & $1$ & $-1$ & $1$ & $1$ & $1$ & $1$ & $-1$ & $1$ & $0.500$ \\
 $-1$ & $-1$ & $1$ & $-1$ & $-1$ & $1$ & $1$ & $1$ & $1$ & $-1$ & $1$ & $0.525$ \\
 $-1$ & $-1$ & $-1$ & $1$ & $-1$ & $1$ & $1$ & $1$ & $1$ & $-1$ & $-1$ & $0.550$ \\
 $-1$ & $-1$ & $1$ & $-1$ & $-1$ & $1$ & $1$ & $1$ & $1$ & $-1$ & $1$ & $0.575$ \\
 $-1$ & $1$ & $1$ & $1$ & $-1$ & $1$ & $1$ & $1$ & $1$ & $-1$ & $-1$ & $0.600$ \\
 $-1$ & $-1$ & $1$ & $-1$ & $-1$ & $1$ & $1$ & $1$ & $1$ & $-1$ & $1$ & $0.625$ \\
 $-1$ & $-1$ & $-1$ & $-1$ & $-1$ & $1$ & $1$ & $1$ & $1$ & $-1$ & $-1$ & $0.650$ \\
 $-1$ & $-1$ & $-1$ & $-1$ & $-1$ & $1$ & $1$ & $1$ & $1$ & $-1$ & $1$ & $0.675$ \\
 $-1$ & $-1$ & $1$ & $-1$ & $-1$ & $1$ & $1$ & $1$ & $1$ & $-1$ & $1$ & $0.700$ \\
 $-1$ & $-1$ & $-1$ & $1$ & $-1$ & $1$ & $-1$ & $1$ & $1$ & $-1$ & $1$ & $0.725$ \\
 $-1$ & $-1$ & $-1$ & $-1$ & $-1$ & $1$ & $1$ & $1$ & $1$ & $-1$ & $-1$ & $0.750$ \\
 $-1$ & $1$ & $-1$ & $1$ & $-1$ & $1$ & $1$ & $1$ & $1$ & $-1$ & $1$ & $0.775$ \\
 $1$ & $1$ & $-1$ & $1$ & $-1$ & $1$ & $1$ & $1$ & $1$ & $-1$ & $1$ & $0.800$ \\
 $-1$ & $-1$ & $1$ & $-1$ & $-1$ & $1$ & $1$ & $1$ & $1$ & $-1$ & $-1$ & $0.825$ \\
 $-1$ & $-1$ & $1$ & $-1$ & $-1$ & $1$ & $1$ & $1$ & $1$ & $-1$ & $-1$ & $0.850$ \\
 $-1$ & $1$ & $1$ & $1$ & $-1$ & $1$ & $1$ & $1$ & $1$ & $-1$ & $-1$ & $0.875$ \\
 $1$ & $1$ & $1$ & $1$ & $1$ & $1$ & $-1$ & $-1$ & $1$ & $1$ & $-1$ & $0.900$ \\
 $-1$ & $1$ & $-1$ & $-1$ & $-1$ & $1$ & $1$ & $1$ & $1$ & $-1$ & $1$ & $0.925$ \\
 $-1$ & $1$ & $-1$ & $1$ & $-1$ & $1$ & $1$ & $1$ & $1$ & $-1$ & $-1$ & $0.950$ \\
 $-1$ & $-1$ & $-1$ & $1$ & $-1$ & $1$ & $1$ & $1$ & $1$ & $-1$ & $-1$ & $0.975$ \\
 $1$ & $-1$ & $1$ & $1$ & $-1$ & $1$ & $1$ & $1$ & $1$ & $-1$ & $-1$ & $1.000$ \\
 $-1$ & $-1$ & $-1$ & $1$ & $-1$ & $1$ & $1$ & $1$ & $1$ & $-1$ & $-1$ & $1.025$ \\
 $-1$ & $1$ & $-1$ & $-1$ & $-1$ & $1$ & $1$ & $1$ & $1$ & $-1$ & $-1$ & $1.050$ \\
 $-1$ & $-1$ & $1$ & $1$ & $-1$ & $1$ & $1$ & $1$ & $1$ & $-1$ & $-1$ & $1.075$ \\
 $-1$ & $-1$ & $-1$ & $-1$ & $-1$ & $1$ & $1$ & $1$ & $1$ & $-1$ & $-1$ & $1.100$ \\
 $-1$ & $-1$ & $-1$ & $-1$ & $-1$ & $1$ & $1$ & $1$ & $1$ & $-1$ & $1$ & $1.125$ \\
 $-1$ & $-1$ & $-1$ & $1$ & $-1$ & $1$ & $1$ & $1$ & $1$ & $-1$ & $1$ & $1.150$ \\
 $-1$ & $1$ & $-1$ & $-1$ & $-1$ & $1$ & $1$ & $1$ & $1$ & $-1$ & $1$ & $1.175$ \\
 $-1$ & $-1$ & $-1$ & $1$ & $-1$ & $1$ & $1$ & $1$ & $1$ & $-1$ & $-1$ & $1.200$ \\
 $1$ & $-1$ & $-1$ & $-1$ & $-1$ & $1$ & $1$ & $1$ & $1$ & $-1$ & $-1$ & $1.225$ \\
 $-1$ & $-1$ & $-1$ & $-1$ & $-1$ & $1$ & $1$ & $1$ & $1$ & $-1$ & $-1$ & $1.250$ \\
 $1$ & $1$ & $-1$ & $1$ & $-1$ & $1$ & $1$ & $1$ & $1$ & $-1$ & $-1$ & $1.275$ \\
 $-1$ & $-1$ & $-1$ & $-1$ & $-1$ & $1$ & $1$ & $1$ & $1$ & $-1$ & $1$ & $1.300$ \\
 $-1$ & $1$ & $-1$ & $-1$ & $-1$ & $1$ & $1$ & $1$ & $1$ & $-1$ & $1$ & $1.325$ \\
 $-1$ & $-1$ & $-1$ & $1$ & $-1$ & $1$ & $1$ & $1$ & $1$ & $-1$ & $1$ & $1.350$ \\
 $-1$ & $-1$ & $-1$ & $1$ & $-1$ & $1$ & $-1$ & $1$ & $1$ & $-1$ & $1$ & $1.375$ \\
 $-1$ & $-1$ & $-1$ & $-1$ & $-1$ & $1$ & $1$ & $1$ & $1$ & $-1$ & $-1$ & $1.400$ \\
 $1$ & $-1$ & $1$ & $-1$ & $-1$ & $1$ & $1$ & $1$ & $1$ & $-1$ & $-1$ & $1.425$ \\
 $1$ & $1$ & $-1$ & $-1$ & $-1$ & $1$ & $1$ & $1$ & $1$ & $-1$ & $1$ & $1.450$ \\
 $-1$ & $1$ & $-1$ & $-1$ & $-1$ & $1$ & $1$ & $1$ & $1$ & $-1$ & $1$ & $1.475$ \\
 $-1$ & $1$ & $-1$ & $-1$ & $-1$ & $1$ & $1$ & $-1$ & $1$ & $-1$ & $-1$ & $1.500$ \\
 $-1$ & $-1$ & $1$ & $1$ & $-1$ & $1$ & $1$ & $1$ & $1$ & $-1$ & $-1$ & $1.525$ \\
 $-1$ & $-1$ & $-1$ & $-1$ & $-1$ & $1$ & $1$ & $1$ & $1$ & $-1$ & $1$ & $1.550$ \\
 $-1$ & $-1$ & $-1$ & $1$ & $-1$ & $1$ & $-1$ & $1$ & $1$ & $-1$ & $1$ & $1.575$ \\
 $-1$ & $1$ & $-1$ & $-1$ & $-1$ & $1$ & $1$ & $1$ & $1$ & $-1$ & $-1$ & $1.600$ \\
 $-1$ & $-1$ & $-1$ & $1$ & $-1$ & $1$ & $1$ & $1$ & $1$ & $-1$ & $1$ & $1.625$ \\
 $-1$ & $-1$ & $-1$ & $1$ & $-1$ & $1$ & $1$ & $1$ & $1$ & $-1$ & $1$ & $1.650$ \\
 $1$ & $1$ & $-1$ & $-1$ & $-1$ & $1$ & $1$ & $1$ & $1$ & $-1$ & $1$ & $1.675$ \\
 $-1$ & $1$ & $1$ & $1$ & $-1$ & $1$ & $1$ & $1$ & $1$ & $-1$ & $-1$ & $1.700$ \\
 $1$ & $1$ & $-1$ & $-1$ & $1$ & $1$ & $-1$ & $-1$ & $1$ & $1$ & $-1$ & $1.725$ \\
 $-1$ & $-1$ & $-1$ & $-1$ & $-1$ & $1$ & $1$ & $1$ & $1$ & $-1$ & $-1$ & $1.750$ \\
 $-1$ & $-1$ & $-1$ & $1$ & $-1$ & $1$ & $1$ & $1$ & $1$ & $-1$ & $-1$ & $1.775$ \\
 $-1$ & $-1$ & $1$ & $-1$ & $-1$ & $1$ & $1$ & $1$ & $1$ & $-1$ & $1$ & $1.800$ \\
 $-1$ & $-1$ & $-1$ & $-1$ & $-1$ & $1$ & $1$ & $1$ & $1$ & $-1$ & $-1$ & $1.825$ \\
 $1$ & $-1$ & $1$ & $-1$ & $1$ & $1$ & $-1$ & $-1$ & $1$ & $1$ & $-1$ & $1.850$ \\
 $1$ & $-1$ & $1$ & $1$ & $-1$ & $1$ & $1$ & $1$ & $1$ & $-1$ & $-1$ & $1.875$ \\
 $-1$ & $-1$ & $-1$ & $-1$ & $-1$ & $1$ & $1$ & $1$ & $1$ & $-1$ & $1$ & $1.900$ \\
 $-1$ & $-1$ & $-1$ & $1$ & $-1$ & $1$ & $1$ & $1$ & $-1$ & $-1$ & $1$ & $1.925$ \\
 $-1$ & $-1$ & $-1$ & $1$ & $-1$ & $1$ & $1$ & $1$ & $1$ & $-1$ & $-1$ & $1.950$ \\
 $-1$ & $1$ & $1$ & $-1$ & $-1$ & $1$ & $1$ & $1$ & $1$ & $-1$ & $1$ & $1.975$ \\
\hline
\end{tabular}
\label{tab:h2e}
\end{table}

\end{document}